\documentclass[preprintnumbers,unsortedaddress,superscriptaddress,notitlepage,square,sort&compress,comma,numbers]{revtex4-1}
\usepackage{eurosym}
\usepackage{amsfonts}
\usepackage{amsmath}
\usepackage{mathtools}
\usepackage{hyperref}
\usepackage{amssymb}
\usepackage[english]{babel}
\usepackage[utf8]{inputenc}
\usepackage{graphicx}
\usepackage{subcaption}
\usepackage{epsfig}
\usepackage{bm}
\usepackage{verbatim}
\usepackage{float}
\usepackage{wrapfig}
\usepackage{booktabs}
\usepackage{natbib}
\usepackage{multirow}
\usepackage[usenames, dvipsnames,x11names]{xcolor}
\usepackage{hyperref}
\usepackage{bm,bbm}
\usepackage{float}
\hypersetup{colorlinks=true, urlcolor=Turquoise4, linkcolor=Aquamarine, citecolor=Orchid4}
  \numberwithin{equation}{section}

\newcommand{\mathsym}[1]{{}}

\input{tcilatex}
\begin{document}

\title{Fermion masses and mixings and some phenomenological aspects of a 3-3-1 model with linear seesaw mechanism}
\author{A. E. C\'{a}rcamo Hern\'{a}ndez}
\email{antonio.carcamo@usm.cl}
\author{Nicol\'{a}s A. P\'{e}rez-Julve}
\email{nicolasperezjulve@gmail.com}
\author{Yocelyne Hidalgo Velásquez}
\email{yocehidalgov@gmail.com}
\affiliation{Universidad T\'{e}cnica Federico Santa Mar\'{\i}a and Centro Cient\'{\i}%
fico-Tecnol\'{o}gico de Valpara\'{\i}so, \\
Casilla 110-V, Valpara\'{\i}so, Chile,}

\date{\today }

\begin{abstract}
We propose a viable theory based on the $SU(3)_C\times SU(3)_L\times U(1)_X$ gauge group supplemented by the $S_4$ discrete group together with other various symmetries, whose spontaneous breaking gives rise to the current SM fermion mass and mixing hierarchy. In the proposed theory the small light active neutrino masses are generated from a linear seesaw mechanism mediated by three Majorana neutrinos. The model is capable of reproducing the experimental values of the physical observables of both quark and lepton sectors. Our model is predictive in the quark sector having 9 effective parameters that allow to successfully reproduce the four CKM parameters and the six Standard Model (SM) quark masses. In the SM quark sector, there is particular scenario, motivated by naturalness arguments, which allows a good fit for its ten observables, with only six effective parameters. We also study the single heavy scalar production via gluon fusion mechanism at proton-proton collider. Our model is also consistent with the experimental constraints arising from the Higgs diphoton decay rate.
\end{abstract}

\maketitle

\section{Introduction}

Although the Standard Model (SM) is a very well established quantum field theory highly consistent with the experimental data, it has several unexplained issues. For instance, the current pattern of SM fermion masses and mixing angles, the number of SM fermion families, the tiny values of active neutrino masses are some of the issues that do not find an explanation within the context of the SM. The SM fermion mass hierarchy is spanned over a range of 13 orders of magnitude from the light
active neutrino mass scale up to the top quark mass. In addition, the
experimental data shows that the quark mixing pattern is significantly
different from the leptonic mixing one. The mixing angles of the quark sector are small, thus implying that the Cabbibo-Kobayashi-Maskawa (CKM) quark mixing matrix is close to the identity matrix. On the other hand, two of
the leptonic mixing angles are large and one is small, of the order of the
Cabbibo angle, thus implying a Pontecorvo-Maki-Nakagawa-Sakata (PMNS) leptonic mixing matrix very different
from the identity matrix. This is the so called flavor puzzle, which is not
addressed by the SM and provides reason for considering models
 with augmented field content and extended symmetry groups added to explain the current
SM fermion mass spectrum and mixing parameters.

Theories with an extended $SU(3)_C\times SU(3)_L\times U(1)_X$ gauge
symmetry \cite%
{Georgi:1978bv,Valle:1983dk,Pisano:1991ee,Foot:1992rh,Frampton:1992wt,Hoang:1996gi, Hoang:1995vq,Foot:1994ym,CarcamoHernandez:2005ka,Dong:2010zu,Dong:2010gk,Dong:2011vb,Benavides:2010zw,Dong:2012bf,Huong:2012pg,Giang:2012vs,Binh:2013axa,Hernandez:2013mcf,Hernandez:2013hea,Hernandez:2014vta,Hernandez:2014lpa,Kelso:2014qka,Vien:2014gza,Phong:2014ofa,Phong:2014yca,Boucenna:2014ela,DeConto:2015eia,Boucenna:2015zwa,Boucenna:2015pav,Benavides:2015afa,Hernandez:2015tna,Hue:2015fbb,Hernandez:2015ywg,Fonseca:2016tbn,Vien:2016tmh,Hernandez:2016eod,Fonseca:2016xsy,Deppisch:2016jzl,Reig:2016ewy,CarcamoHernandez:2017cwi,CarcamoHernandez:2017kra,Hati:2017aez,Barreto:2017xix,CarcamoHernandez:2018iel,Vien:2018otl,Dias:2018ddy,Ferreira:2019qpf,CarcamoHernandez:2019vih,CarcamoHernandez:2019lhv}
(3-3-1 models) are used to explain the origin of the three family structure
in the fermion sector, which is left unexplained in the SM. In these models,
the chiral anomaly cancellation condition is fulfilled when there are equal
number of $SU(3)_L$ fermionic triplets and antitriplets, which occurs when
the number of fermion families is a multiple of three. In addition, when the
chiral anomaly cancellation condition is combined with the asymptotic
freedom in QCD, theories based on the $SU(3)_C\times SU(3)_L\times U(1)_X$
gauge symmetry predict the existence of three fermion families. Furthermore,
the large mass difference between the heaviest quark and the two lighter
ones can be explained in 3-3-1 models due to the fact that the third family
is treated under a different representation than the first and second ones. 
 
Furthermore, the 3-3-1 models explain the quantization of the electric
charge \cite{deSousaPires:1998jc,VanDong:2005ux}, have sources of CP
violation \cite{Montero:1998yw,Montero:2005yb}, have a natural Peccei-Quinn
symmetry, which solves the strong-CP problem \cite%
{Pal:1994ba,Dias:2002gg,Dias:2003zt,Dias:2003iq}, predict the limit $%
\sin\theta^2_W<\frac1{4}$, for the weak mixing parameter. Besides that, if
one includes heavy sterile neutrinos in the fermionic spectrum of the 3-3-1
models, such theories will have cold dark matter candidates as weakly
interacting massive particles (WIMPs) \cite%
{Mizukoshi:2010ky,Dias:2010vt,Alvares:2012qv,Cogollo:2014jia}. A concise
review of WIMPs in 3-3-1 Electroweak Gauge Models is provided in Ref. \cite%
{daSilva:2014qba}. Finally, if one considers 3-3-1 Electroweak Gauge Models
with three right handed Majorana neutrinos and without exotic charges, one
can implement a low scale linear or inverse seesaw mechanism, useful for
generating the tiny active neutrinos masses.

In this work, motivated by the aforementioned considerations, we propose an
extension of the 3-3-1 model with right handed Majorana neutrinos, where the
scalar spectrum is enlarged by the inclusion of several gauge singlet
scalars. Our theoretical construction successfully explains the current SM fermion mass spectrum
and fermionic mixing parameters. In the proposed model, the $SU(3)_C\times
SU(3)_L\times U(1)_X$ gauge symmetry is supplemented by the $S_4$ family
symmetry and other auxiliary cyclic symmetries, whose spontaneous breaking
produces the current SM fermion mass spectrum and mixing parameters. In the proposed model, the masses for the Standard Model charged fermions lighter than the top quark are produced by a Froggatt-Nielsen mechanism and the tiny
masses for the light active neutrinos are generated by a linear seesaw
mechanism. We employ the $S_4$ family symmetry because it is the smallest non abelian
group having a doublet, triplet and singlet irreducible representations,
thus permitting to accommodate the three fermion families of the SM.
It is worth mentioning that the $S_4$ discrete group \cite%
{Altarelli:2009gn,Bazzocchi:2009da,Bazzocchi:2009pv,Toorop:2010yh,Patel:2010hr,Morisi:2011pm,Altarelli:2012bn,Mohapatra:2012tb,BhupalDev:2012nm,Varzielas:2012pa,Ding:2013hpa,Ishimori:2010fs,Ding:2013eca,Hagedorn:2011un,Campos:2014zaa,Dong:2010zu,VanVien:2015xha,deAnda:2017yeb,deAnda:2018oik,CarcamoHernandez:2019eme,Chen:2019oey,deMedeirosVarzielas:2019cyj,deMedeirosVarzielas:2019hur} has been shown to provide a nice description for the observed pattern of SM
fermion masses and mixing angles.  

The layout of the remainder of the paper is as follows. In section \ref%
{model} we describe the proposed model, its symmetries, particle content and
Yukawa interactions. The gauge sector of the model is described in section %
\ref{gaugesector}, whereas its low energy scalar potential is presented in
section \ref{scalarpotential}. In section \ref{quarksector} we discuss the
implications of our model in quark masses and mixings. In Section %
\ref{leptonsector}, we present our results on lepton masses and mixing. 
The consequences of our model in the Higgs diphoton decay rate are discussed in
section \ref{Higgsdiphotonrate}. The production of the heavy $H_1$ scalar at proton-proton collider is discussed in Section \ref{H1lhc}. We conclude in section \ref{conclusions}. Appendix %
\ref{S4} provides a description of the $S_4$ discrete group. Appendices \ref%
{scalarpotentialS4doublet} and \ref{scalarpotentialS4triplet} present a
discussion of the scalar potentials for a $S_4$ scalar doublet and $S_4$
triplet, respectively.

\section{The model.}

\subsection{Particle spectrum and symmetries}

\label{model} 
We propose an extension of the 3-3-1 model with right handed
Majorana neutrinos, where the $SU(3)_{C}\times SU(3)_{L}\times U(1)_{X}$
gauge symmetry is augmented by the $S_{4}\times Z_{6}\times Z_{12}\times
Z_{16}$ discrete group and the scalar spectrum is enlarged by considering gauge singlet scalars, which are added in order to generate viable
textures for the fermion sector that successfully explain the current pattern of
SM fermion masses and mixing angles. The scalar and fermionic content with
their assignments under the $SU(3)_{C}\times SU(3)_{L}\times U(1)_{X}\times
S_{4}\times Z_{6}\times Z_{12}\times Z_{16}$ group are given in Tables \ref%
{tab:scalars} and \ref{tab:fermions}, respectively. The dimensions of the $%
SU(3)_{C}$, $SU(3)_{L}$ and $S_{4}$ representations shown in Tables \ref%
{tab:scalars} and \ref{tab:fermions}, are described by numbers in
boldface and the additive notation is used to specify the $U(1)_X$ and $Z_{N}$ charges. We choose the $S_{4}$ symmetry since it is the smallest non abelian group having doublet, triplet and singlet irreducible representations, thus allowing us to naturally accommodate the three families of the SM left handed leptonic fields into a $S_{4}$ triplet, the three gauge singlet right handed Majorana neutrinos into one $S_{4}$ singlet and one $S_{4}$ doublet, the three right handed SM down type quarks into a $S_{4}$ singlet and a $S_{4}$ doublet and the remaining fermionic fields as $S_{4}$ singlets. In addition, the $S_4$, $Z_{6}$ and $Z_{12}$ symmetries shape the textures of the SM fermion mass matrices thus yielding a reduction of the model parameters, especially in the SM quark sector. In addition, the $Z_{6}$ and $Z_{12}$ symmetries separate the $S_4$ scalar triplets ($\Phi$, $\Xi$, $\Omega$, $\varphi$, $\phi$) participating in the charged lepton Yukawa interactions, from the ones ($\zeta$, $\Sigma$, $S$) appearing in the neutrino Yukawa terms. Moreover, the $Z_{6}$ symmetry allows to distinguish the $S_4$ scalar triplets $\Phi$, $\Xi$, $\Omega$ generating the third column of the SM charged lepton mass matrix from the ones $\varphi$ and $\phi$ that give rise to the second and the first column of the SM charged lepton mass matrix, respectively. Consequently, the $Z_{6}$ symmetry selects the allowed entries of the SM charged lepton mass matrix, thus allowing to reduce the number of lepton sector model parameters. Furthermore, the $Z_{12}$ symmetry distinguishes the $S_4$ scalar triplet $\zeta$ participating in the Dirac Yukawa interactions, from the ones $\Sigma$ and $S$ appearing in the remaining neutrino Yukawa terms. Besides that, the $Z_{12}$ symmetry also distinguishes the different $S_4$ scalar doublets $\xi$, $\Delta$ and $\Theta$ that contribute to the first, second and third rows of the SM down type quark mass matrix, respectively, thus allowing to obtain a predictive texture for the down type quark sector that generates the SM down quark masses, the Cabbibo mixing as well as the mixing between the second and third quark families. Furthermore, the $Z_{12}$ symmetry also determines the allowed entries of the SM up type quark mass matrix. It is worth mentioning that due to the $Z_{12}$ charge assignments, the only non vanishing entries of the SM up type quark mass matrix are the diagonal ones as well as the $13$ entry, needed to generate the SM up quark masses as well as the quark mixing angle in the 13 plane and the quark CP violating phase, respectively. The quark mixing angle in the 13 plane and the quark CP violating phase only arise from the up type quark sector. 
 The $Z_{16}$ symmetry shapes the hierarchical structure of the SM fermion mass matrices crucial to yield the observed SM fermion mass and mixing pattern. We remark that $Z_{16}$ is the smallest discrete symmetry permitting to build the Yukawa terms
$\left( \overline{l}_L \rho \Pi\right)_{\mathbf{\mathbf{1}}}e_{1R}\frac{\sigma_{2}^8}{\Lambda^9 }$ ($\Pi=\Phi,\Xi,\Omega$) and $\overline{q}_{1L}\rho ^{\ast }u_{1R}\frac{\sigma _{2}^{8}}{\Lambda
^{8}}$, required to provide a natural explanation for the small values
of the electron and up quark masses, which are $\lambda^9%
\frac{v}{\sqrt{2}}$ and $\lambda^8%
\frac{v}{\sqrt{2}}$ times a $\mathcal{O}(1)$ coupling, respectively, where $\lambda=0.225$ is one of the Wolfenstein parameters. In our model, the masses of the Standard Model charged fermions lighter than the top quark arise from a Froggatt-Nielsen mechanism \cite{Froggatt:1978nt}, triggered by non renormalizable Yukawa interactions invariant under the different discrete group factors. Thus, the current pattern of SM fermion masses and mixing angles arises from the $S_{4}\times Z_{6}\times Z_{12}\times Z_{16}$ spontaneous symmetry breaking. The masses of the light active neutrinos are generated from a linear seesaw mechanism \cite{Mohapatra:1986bd,Akhmedov:1995ip,Akhmedov:1995vm,Malinsky:2005bi,Borah:2018nvu,Hirsch:2009mx,Dib:2014fua,Chakraborty:2014hfa,Sinha:2015ooa,Borah:2018nvu}, which can be
implemented in our model because the third component of the $SU\left(3\right) _{L}$ leptonic triplet is electrically neutral and the fermionic
spectrum includes three right handed Majorana neutrinos. In addition, the
non SM fermions in our model do not have exotic electric charges.
Consequently, the electric charge is defined as: 
\begin{equation}
Q=T_{3}+\beta T_{8}+XI=T_{3}-\frac{1}{\sqrt{3}}T_{8}+XI,
\end{equation}%
with $I=diag(1,1,1)$, $T_{3}=\frac{1}{2}diag(1,-1,0)$ and $T_{8}=(\frac{1}{2%
\sqrt{3}})diag(1,1,-2)$ for a $SU(3)_{L}$ triplet.
In our model the full symmetry $\mathcal{G}$ experiences the three-step spontaneous breaking pattern: 
\begin{eqnarray}
&&\mathcal{G}=SU(3)_{C}\times SU\left( 3\right) _{L}\times U\left( 1\right)
_{X}\times S_{4}\times Z_{6}\times Z_{12}\times Z_{16}{\xrightarrow{%
\Lambda_{int}}} \notag \\
&&\hspace{7mm}SU(3)_{C}\times SU\left( 3\right) _{L}\times U\left( 1\right)
_{X}{\xrightarrow{v_{\chi}}}  \notag \\
&&\hspace{7mm}SU(3)_{C}\times SU\left( 2\right) _{L}\times U\left( 1\right)
_{Y}{\xrightarrow{v_{\eta },v_{\rho}}}  \notag \\
&&\hspace{7mm}SU(3)_{C}\times U\left( 1\right) _{Q},  \label{SB}
\end{eqnarray}
where the different symmetry breaking scales satisfy $\Lambda _{int}\gg v_{\chi }\gg v_{\eta },v_{\rho }$, with $\sqrt{v_{\eta
}^{2}+v_{\rho }^{2}}=246$ GeV. The first step of spontaneous symmetry
breaking is produced by all gauge singlet scalar fields (excepting $\zeta $),
charged under the discrete symmetries, assumed to acquire vacuum expectation
values (VEVs) at a very large energy scale $\Lambda _{int}\gg v_{\chi }\sim 
\mathcal{O}(10)$ TeV. The second step of spontaneous symmetry breaking is
caused by the $SU(3)_{L}$ scalar triplet $\chi $, whose third component
acquires a $10$ TeV scale vacuum expectation value (VEV) that breaks the $%
SU(3)_{L}\times U(1)_{X}$ gauge symmetry, thus providing masses for the
exotic fermions, non Standard Model gauge bosons and the heavy CP even
neutral scalar state of $\chi $. We further assume that the $S_{4}$ triplet
gauge singlet scalar $\zeta $ acquires a VEVs at the same scale of $v_{\chi
} $. Finally, the remaining two $SU(3)_{L}$ scalar triplets $\eta $ and $%
\rho $, whose first and second components, respectively, get VEVs at the
Fermi scale, thus producing the masses for the SM particles and for the
physical neutral scalar states arising from those scalar triplets. Here we
are considering that the $SU(3)_{L}\times U(1)_{X}$ gauge symmetry is
spontaneously broken at a scale of about $10$ TeV in order to comply with
collider constraints \cite{Salazar:2015gxa} as well as with the constraints
arising from the experimental data on $K$, $D$ and $B$ meson mixings \cite%
{Huyen:2012uk} and from the $B_{s,d}\rightarrow \mu ^{+}\mu ^{-}$ and $%
B_{d}\rightarrow K^{\ast }(K)\mu ^{+}\mu ^{-}$ decays \cite%
{CarcamoHernandez:2005ka,Martinez:2008jj,Buras:2013dea,Buras:2014yna,Buras:2012dp}%
.

The $SU(3)_{L}$ triplet scalars $\chi $, $\eta $ and $\rho $ can be represented as: 
\begin{equation}
\chi =%
\begin{pmatrix}
\chi _{1}^{0} \\ 
\chi _{2}^{-} \\ 
\frac{1}{\sqrt{2}}(v_{\chi }+\xi _{\chi }\pm i\zeta _{\chi })%
\end{pmatrix}%
,\hspace{1cm}\eta =%
\begin{pmatrix}
\frac{1}{\sqrt{2}}(v_{\eta }+\xi _{\eta }\pm i\zeta _{\eta }) \\ 
\eta _{2}^{-} \\ 
\eta _{3}^{0}%
\end{pmatrix}%
,\hspace{1cm}\rho =%
\begin{pmatrix}
\rho _{1}^{+} \\ 
\frac{1}{\sqrt{2}}(v_{\rho }+\xi _{\rho }\pm i\zeta _{\rho }) \\ 
\rho _{3}^{+}%
\end{pmatrix}%
.
\end{equation}

The $SU(3)_{L}$ fermionic antitriplets and triplets are expressed as: 
\begin{equation}
q_{nL}=%
\begin{pmatrix}
d_{n} \\ 
-u_{n} \\ 
j_{n} \\ 
\end{pmatrix}%
_{L},\hspace{1cm}q_{3L}=%
\begin{pmatrix}
u_{3} \\ 
d_{3} \\ 
t^{\prime} \\ 
\end{pmatrix}%
_{L},\hspace{1cm}L_{iL}=%
\begin{pmatrix}
\nu _{i} \\ 
e_{i} \\ 
\nu _{i}^{c} \\ 
\end{pmatrix}%
_{L},\hspace{1cm}n=1,2,\hspace{1cm}i=1,2,3.
\end{equation}

\begin{table}[th]
\begin{tabular}{|c|c|c|c|c|c|c|c|c|c|c|c|c|c|c|c|c|}
\hline
& $\chi $ & $\eta $ & $\rho $ & $\sigma _{1}$ & $\sigma _{2}$ & $\xi $ & $%
\Delta $ & $\Theta $ & $\Phi $ & $\Xi $ & $\Omega $ & $\varphi$ & $\phi$ & $\zeta $ & $\Sigma $ & 
$S$ \\ \hline
$SU(3)_{C}$ & $\mathbf{1}$ & $\mathbf{1}$ & $\mathbf{1}$ & $\mathbf{1}$ & $%
\mathbf{1}$ & $\mathbf{1}$ & $\mathbf{1}$ & $\mathbf{1}$ & $\mathbf{1}$ & $%
\mathbf{1}$ & $\mathbf{1}$ & $\mathbf{1}$ & $\mathbf{1}$ & $\mathbf{1}$ & $\mathbf{1}$ & $\mathbf{1}$ \\ 
\hline
$SU(3)_{L}$ & $\mathbf{3}$ & $\mathbf{3}$ & $\mathbf{3}$ & $\mathbf{1}$ & $%
\mathbf{1}$ & $\mathbf{1}$ & $\mathbf{1}$ & $\mathbf{1}$ & $\mathbf{1}$ & $%
\mathbf{1}$ & $\mathbf{1}$ & $\mathbf{1}$ & $\mathbf{1}$ & $\mathbf{1}$ & $\mathbf{1}$ & $\mathbf{1}$ \\ 
\hline
$U(1)_{X}$ & $-\frac{1}{3}$ & $-\frac{1}{3}$ & $\frac{2}{3}$ & $0$ & $0$ & $%
0 $ & $0$ & $0$ & $0$ & $0$ & $0$ & $0$ & $0$ & $0$ & $0$ & $0$ \\ \hline
$S_{4}$ & $\mathbf{1}$ & $\mathbf{1}$ & $\mathbf{1}$ & $\mathbf{1}$ & $%
\mathbf{1}^{\prime }$ & $\mathbf{2}$ & $\mathbf{2}$ & $\mathbf{2}$ & $%
\mathbf{3}$ & $\mathbf{3}$ & $\mathbf{3}$ & $\mathbf{3}$ & $\mathbf{3}^{\prime }$ & $\mathbf{3}$ & $\mathbf{3}%
^{\prime }$ & $\mathbf{3}^{\prime }$ \\ \hline
$Z_{6}$ & $0$ & $0$ & $0$ & $0$ & $0$ & $0$ & $0$ & $0$ & $2$ & $2$ & $2$ & $1$ & $1$ & $0$ & $0$ & $0$ \\ \hline
$Z_{12}$ & $0$ & $0$ & $0$ & $-3$ & $1$ & $-5$ & $-4$ & $-2$ & $1$ & $0$ & $1$ & $1$ & $1$ & $-2$ & $0$ & $0$ \\ \hline
$Z_{16}$ & $0$ & $0$ & $0$ & $-1$ & $-1$ & $-2$ & $-1$ & $-1$ & $0$ & $1$ & $0$ & $0$ & $0$ & $0$ & $0$ & $1$ \\ \hline
\end{tabular}%
\caption{Scalar transformations under the $SU(3)_{C}\times SU(3)_{L}\times
U(1)_{X}\times S_{4}\times Z_{6}\times Z_{12}\times Z_{16}$ group.}
\label{tab:scalars}
\end{table}
\begin{table}[th]
\begin{tabular}{|c|c|c|c|c|c|c|c|c|c|c|c|c|c|c|c|c|c|}
\hline
& $q_{1L}$ & $q_{2L}$ & $q_{3L}$ & $u_{1R}$ & $u_{2R}$ & $u_{3R}$ & $d_{1R}$
& $d_{R}$ & $t^{\prime}_{R}$ & $j_{1R}$ & $j_{2R}$ & $L_{L}$ & $e_{1R}$ & $e_{2R}$ & $%
e_{3R}$ & $N_{1R}$ & $N_{R}$ \\ \hline
$SU(3)_{C}$ & $\mathbf{3}$ & $\mathbf{3}$ & $\mathbf{3}$ & $\mathbf{3}$ & $%
\mathbf{3}$ & $\mathbf{3}$ & $\mathbf{3}$ & $\mathbf{3}$ & $\mathbf{3}$ & $%
\mathbf{3}$ & $\mathbf{3}$ & $\mathbf{1}$ & $\mathbf{1}$ & $\mathbf{1}$ & $%
\mathbf{1}$ & $\mathbf{1}$ & $\mathbf{1}$ \\ \hline
$SU(3)_{L}$ & $\mathbf{3^{\ast }}$ & $\mathbf{3^{\ast }}$ & $\mathbf{3}$ & $%
\mathbf{1} $ & $\mathbf{1}$ & $\mathbf{1}$ & $\mathbf{1}$ & $\mathbf{1}$ & $%
\mathbf{1}$ & $\mathbf{1}$ & $\mathbf{1}$ & $\mathbf{3}$ & $\mathbf{1}$ & $%
\mathbf{1}$ & $\mathbf{1}$ & $\mathbf{1}$ & $\mathbf{1}$ \\ \hline
$U(1)_{X}$ & $0$ & $0$ & $\frac{1}{3}$ & $\frac{2}{3}$ & $\frac{2}{3}$ & $%
\frac{2}{3}$ & $-\frac{1}{3}$ & $-\frac{1}{3}$ & $\frac{2}{3}$ & $-\frac{1}{3%
}$ & $-\frac{1}{3}$ & $-\frac{1}{3}$ & $-1$ & $-1$ & $-1$ & $0$ & $0$ \\ 
\hline
$S_{4}$ & $\mathbf{1}$ & $\mathbf{1}$ & $\mathbf{1}$ & $\mathbf{1}$ & $%
\mathbf{1}$ & $\mathbf{1}$ & $\mathbf{1}^{\prime }$ & $\mathbf{2}$ & $%
\mathbf{1}$ & $\mathbf{1}$ & $\mathbf{1}$ & $\mathbf{3}$ & $\mathbf{1}^{\prime }$ & $%
\mathbf{1}$ & $\mathbf{1}$ & $\mathbf{1}^{\prime }$ & $\mathbf{2}$ \\ \hline
$Z_{6}$ & $0$ & $0$ & $0$ & $0$ & $0$ & $0$ & $0$ & $0$ & $0$ & $0$ & $0$ & 
$0$ & $-1$ & $-1$ & $-2$ & $0$ & $0$ \\ \hline
$Z_{12}$ & $0$ & $0$ & $0$ & $4$ & $0$ & $0$ & $5$ & $0$ & $0$ & $0$ & $0$ & 
$-1$ & $2$ & $-6$ & $-4$ & $-1$ & $-1$ \\ \hline
$Z_{16}$ & $-4$ & $-2$ & $0$ & $4$ & $2$ & $0$ & $3$ & $3$ & $0$ & $-4$ & $-2$
& $0$ & $8$ & $4$ & $2$ & $0$ & $-1$ \\ \hline
\end{tabular}%
\caption{Fermion transformations under the $SU(3)_{C}\times SU(3)_{L}\times
U(1)_{X}\times S_{4}\times Z_{6}\times Z_{12}\times Z_{16}$  group.}
\label{tab:fermions}
\end{table}
Using the particle spectrum and symmetries given in Tables \ref{tab:scalars} and \ref%
{tab:fermions}, we can write the Yukawa interactions for the quark and lepton sectors: 
\begin{eqnarray}
-\mathcal{L}_{Y}^{\left( q\right) } &=&y^{\left( t^{\prime}\right) }\overline{q}%
_{3L}\chi t^{\prime}_{R}+y_{33}^{\left( u\right) }\overline{q}_{3L}\eta
u_{3R}+y_{13}^{\left( u\right) }\overline{q}_{1L}\rho ^{\ast }u_{3R}\frac{%
\sigma _{1}^{4}}{\Lambda ^{4}}+y_{22}^{\left( u\right) }\overline{q}%
_{2L}\rho ^{\ast }u_{2R}\frac{\sigma _{1}^{4}}{\Lambda ^{4}}+y_{11}^{\left(
u\right) }\overline{q}_{1L}\rho ^{\ast }u_{1R}\frac{\sigma _{2}^{8}}{\Lambda
^{8}}  \notag \\
&&+y_{1}^{\left( j\right) }\overline{q}_{1L}\chi ^{\ast
}j_{1R}+y_{2}^{\left( j\right) }\overline{q}_{2L}\chi ^{\ast
}j_{2R}+y_{3}^{\left( d\right) }\overline{q}_{3L}\rho \left( \Theta
d_{R}\right) _{\mathbf{\mathbf{1}}}\frac{\sigma _{2}^{2}}{\Lambda ^{3}}%
+y_{2}^{\left( d\right) }\overline{q}_{2L}\eta ^{\ast }\left( \Delta
d_{R}\right) _{\mathbf{\mathbf{1}}}\frac{\sigma _{2}^{4}}{\Lambda ^{5}}%
+y_{4}^{\left( d\right) }\overline{q}_{1L}\eta ^{\ast }\left( \xi
d_{R}\right) _{\mathbf{\mathbf{1}}}\frac{\sigma _{2}^{5}}{\Lambda ^{6}} 
\notag \\
&&+y_{1}^{\left( d\right) }\overline{q}_{1L}\eta ^{\ast }d_{1R}\frac{\sigma
_{2}^{7}}{\Lambda ^{7}}+H.c,  \label{Lyq}
\end{eqnarray}%
\begin{eqnarray}
-\mathcal{L}_{Y}^{\left( l\right) } &=&x_{1}^{\left( L\right) }\left( 
\overline{L}_{L}\rho \phi \right) _{\mathbf{\mathbf{1}}}e_{1R}\frac{\sigma
_{2}^{8}}{\Lambda ^{9}}+y_{1}^{\left( L\right) }\left( \overline{L}_{L}\rho \varphi \right) _{%
\mathbf{\mathbf{1}}}e_{2R}\frac{\sigma _{2}^{4}}{\Lambda ^{5}}+y_{2}^{\left(
L\right) }\left( \overline{L}_{L}\rho \varphi \right) _{\mathbf{\mathbf{1}}%
}e_{2R}\frac{\sigma _{2}^{4}}{\Lambda ^{5}}\notag \\
&&+z_{1}^{\left( L\right) }\left( \overline{L}_{L}\rho \Phi \right) _{%
\mathbf{\mathbf{1}}}e_{3R}\frac{\sigma _{2}^{2}}{\Lambda ^{3}}+z_{2}^{\left(
L\right) }\left( \overline{L}_{L}\rho \Xi \right) _{\mathbf{\mathbf{1}}%
}e_{3R}\frac{\sigma_{2}^{3}}{\Lambda^{4}}+z_{3}^{\left( L\right) }\left( 
\overline{L}_{L}\rho \Omega \right) _{\mathbf{\mathbf{1}}}e_{3R}\frac{\sigma
_{2}^{2}}{\Lambda ^{3}}+y_{\rho }\varepsilon _{abc}\left( \overline{L}%
_{L}^{a}\left( L_{L}^{C}\right) ^{b}\right) _{\mathbf{3}}\left( \rho ^{\ast
}\right) ^{c}\frac{\zeta }{\Lambda }  \notag \\
&&+y_{1\chi }^{\left( L\right) }\left( \overline{L}_{L}\Sigma \right) _{%
\mathbf{1}^{\prime }}\chi N_{1R}\frac{1}{\Lambda }+y_{2\chi }^{\left(
L\right) }\overline{L}_{L}\chi \left( SN_{R}\right) _{\mathbf{3}}\frac{1}{%
\Lambda }+y_{1\eta }^{\left( L\right) }\left( \overline{L}_{L}\Sigma \right)
_{\mathbf{1}^{\prime }}\eta N_{1R}\frac{1}{\Lambda }+y_{2\eta }^{\left(
L\right) }\overline{L}_{L}\eta \left( SN_{R}\right) _{\mathbf{3}}\frac{1}{%
\Lambda }+H.c  \label{Lyl}
\end{eqnarray}
 As shown in detail in Appendices \ref{scalarpotentialS4doublet} and \ref{scalarpotentialS4triplet}, the following VEV configurations for the $S_{4}$
 scalar doublets and $S_{4}$ scalar triplets are in accordance with the minimization conditions of the scalar potential:
\begin{eqnarray}
\left\langle \xi \right\rangle &=&v_{\xi }\left( 0,-1\right) ,\hspace{1cm}%
\left\langle \Delta \right\rangle =\frac{v_{\Delta }}{\sqrt{5}}\left(
2,1\right) ,\hspace{1cm}\left\langle \Theta \right\rangle =v_{\Theta }\left(
1,0\right) ,\hspace{1cm}\left\langle \Phi \right\rangle =v_{\Phi }\left(
1,0,0\right) ,\hspace{1cm}\left\langle \Xi \right\rangle =\frac{v_{\Xi }}{%
\sqrt{2}}\left(0,1,0\right),\notag \\
\left\langle \Omega \right\rangle &=&\frac{v_{\Omega }}{\sqrt{2}}\left(
0,0,1\right),\hspace{1cm}\left\langle\varphi\right\rangle=\frac{v_{\varphi}%
}{\sqrt{1+r^{2}}}\left( 1,0,r\right),\hspace{1cm}\left\langle\phi\right\rangle=v_{\phi}\left(1,0,0\right),\notag \\
\left\langle \zeta \right\rangle &=&\frac{v_{\zeta }%
}{\sqrt{1+c^{2}}}\left( 1,0,c\right) ,\hspace{1cm}\left\langle \Sigma
\right\rangle =\frac{v_{\Sigma }}{\sqrt{3}}\left( 1,-1,1\right) ,\hspace{1cm}%
\left\langle S\right\rangle =\frac{v_{S}}{\sqrt{3}}\left( 1,1,1\right) .
\label{eqn:VEV}
\end{eqnarray}
Given that the spontaneous $S_{4}\times Z_{6}\times Z_{12}\times
Z_{16}$ symmetry breaking gives rise to the observed hierarchy of SM charged fermion masses and quark
mixing parameters, the vacuum expectation values of the gauge singlet scalars can be expressed 
in terms of the Wolfenstein parameter $\lambda =0.225$ and the model
cutoff $\Lambda $, in the following way:
\begin{equation}
v_{\eta }\sim v_{\rho }<<v_{\zeta }\sim v_{\chi }<<v_{\Delta }\sim v_{\Phi
}\sim v_{\Xi }\sim v_{\Omega }\sim v_{\xi }\sim v_{\Theta }\sim v_{\Sigma
}\sim v_{S}\sim v_{\sigma _{n}}\sim v_{\varphi}\sim v_{\phi}\sim \Lambda _{int}=\lambda \Lambda ,\hspace{%
0.5cm}n=1,2.\label{VEVsinglets}
\end{equation}
where the model cutoff $\Lambda $ can be interpreted as the scale of the UV completion of the model, e.g. the masses of Froggatt-Nielsen messenger fields.

\subsection{The gauge sector}\label{gaugesector}

Using $\beta = -1/\sqrt{3}$, the gauge bosons related with $SU(3)_L$ are:
\begin{align}
\mathbf{W}_\mu &= W^{\alpha}_{\mu}G_{\alpha}  \notag \\
&= \frac{1}{2} 
\begin{pmatrix}
W^{3}_{\mu} + \frac{1}{\sqrt{3}}W^{8}_{\mu} & \sqrt{2}W^{+}_{\mu} & \sqrt{2}%
K^{0}_{\mu} \\ 
\sqrt{2}W^{-}_{\mu} & -W^{3}_{\mu} + W^{8}_{\mu} & \sqrt{2}K^{-}_{\mu} \\ 
\sqrt{2}\overline{K}^{0}_{\mu} & \sqrt{2}K^{+}_{\mu} & -\frac{2}{\sqrt{3}}%
W^{8}_{\mu}%
\end{pmatrix}
\label{eqn:w}
\end{align}
where $G_{\alpha}$ ($\alpha=1,\cdots,8$) are the Gell-Mann matrices. The representation of the gauge field related to the $U(1)_X$ gauge symmetry has $Q_{B}=0$ charge and is given by: 
\begin{align}
\mathbf{B}_{\mu} = \mathbf{I}_{3\times3}B_{\mu}
\end{align}
In this model three gauge fields with no electric charge are combined to form the photon as well as the $Z$ and $Z^{\prime }$ gauge bosons. Moreover, there are two gauge fields with electric charge $\pm 1$ (%
$W^{\pm}$) and four non SM gauge bosons $W^{\prime \pm}$, $\overline{K}^{0}$, $K^{0}$.

The covariant derivative in 3-3-1 models reads: 
\begin{equation}
D_{\mu} = \partial_{\mu} + igW^{\alpha}_{\mu}G_{\alpha} + ig^{\prime
}X_{\Phi}B_{\mu}  \label{eqn:covder}
\end{equation}

Replacing Eq. (\ref{eqn:covder}) in the scalar kinetic interactions gives rise to the gauge boson mass terms as well as to the interactions between the scalar and gauge bosons \cite{Diaz:2003dk}: 
\begin{align}
\mathcal{L}_{Kin}=&\sum_{\Phi=\eta,\rho,\chi} (D^{\mu}\Phi)^{\dagger}(D_{\mu}\Phi)  \notag \\
=&\sum_{\Phi=\eta,\rho,\chi}\left[\overbrace{(\partial^{\mu}\Phi)^{\dagger}(D_{\mu}\Phi)+(D^{\mu}\Phi)^{%
\dagger}(\partial_{\mu}\Phi)}^{(1)}-(\partial^{\mu}\Phi)^{\dagger}(%
\partial_{\mu}\Phi) + \overbrace{\Phi^{\dagger}(gW^{\mu}+g^{\prime
}X_{\Phi}B^{\mu})^{\dagger}(gW^{\mu}+g^{\prime }X_{\Phi}B^{\mu})\Phi}^{(2)}\right],
\label{eqn:lagrangian}
\end{align}
where the terms denoted as $(1)$ give rise to the interactions between the Goldstone and massive gauge bosons. On the other hand, the terms denoted as $(2)$ produce the masses for the gauge bosons and the interactions between the gauge bosons and the physical scalar fields. The gauge boson squared mass matrices read:
\begin{gather}
    M^{2}_{charged} = \left(
\begin{array}{cc}
 \frac{1}{8} g^2 v _{\text{$\eta $}}^2+\frac{1}{8} g^2 v
   _{\text{$\rho $}}^2 & 0 \\
 0 & \frac{1}{8} g^2 v _{\text{$\rho $}}^2+\frac{1}{8} g^2
   v _{\text{$\chi $}}^2 \\
\end{array}
\right)\\
M^{2}_{neutral} =
\left(
\begin{array}{cccc}
 \frac{1}{4} g^2 (v_{\text{$\eta $}}^2+ v
   _{\text{$\rho$}}^2) & \frac{g^2 v _{\text{$\eta $}}^2}{4
   \sqrt{3}}-\frac{g^2 v _{\text{$\rho $}}^2}{4 \sqrt{3}} &
   -\frac{1}{6} g g' v _{\text{$\eta $}}^2-\frac{1}{3} g v
   _{\text{$\rho $}}^2 g' & 0 \\
 \frac{g^2 v _{\text{$\eta $}}^2}{4 \sqrt{3}}-\frac{g^2 v
   _{\text{$\rho $}}^2}{4 \sqrt{3}} & \frac{1}{12} g^2 v
   _{\text{$\eta $}}^2+\frac{1}{12} g^2 v _{\text{$\rho
   $}}^2+\frac{1}{3} g^2 v _{\text{$\chi $}}^2 & -\frac{g
   g' v _{\text{$\eta $}}^2}{6 \sqrt{3}}+\frac{g v
   _{\text{$\rho $}}^2 g'}{3 \sqrt{3}}+\frac{g v
   _{\text{$\chi $}}^2 g'}{3 \sqrt{3}} & 0 \\
 -\frac{1}{6} g g' v _{\text{$\eta $}}^2-\frac{1}{3} g v
   _{\text{$\rho $}}^2 g' & -\frac{g g' v _{\text{$\eta
   $}}^2}{6 \sqrt{3}}+\frac{g v _{\text{$\rho $}}^2 g'}{3
   \sqrt{3}}+\frac{g v _{\text{$\chi $}}^2 g'}{3 \sqrt{3}} &
   \frac{1}{9} v _{\text{$\eta $}}^2
   \left(g'\right)^2+\frac{4}{9} v _{\text{$\rho $}}^2
   \left(g'\right)^2+\frac{1}{9} v _{\text{$\chi $}}^2
   \left(g'\right)^2 & 0 \\
 0 & 0 & 0 & \frac{1}{8} g^2 (v _{\text{$\eta
   $}}^2 + v _{\text{$\chi $}}^2) \\
\end{array}
\right)
\end{gather}

The physical gauge bosons and their masses are shown in Table \ref{tab:gaugebosons}:

\begin{minipage}{0.45\textwidth}
\begin{table}[H]
\centering
\renewcommand{\arraystretch}{1.4}
\begin{tabular}{c|c}
\textbf{Gauge Boson} & \textbf{Square Mass}  \\
\hline\hline\
$W^{\pm}$ & $\frac{1}{4}g^2 (v_{\eta}^2 + v^{2}_{\rho})$ \\
\hline
$W'^{\pm}$ & $\frac{1}{4}g^2 (v_{\chi}^2 + v^{2}_{\rho})$ \\
\hline
$\gamma$ & 0 \\ \hline
$Z$ & $\frac{1}{9} \left( \Xi_1 - \Xi_2 \right)$  \\ 
\hline
$Z^{\prime }$ & $\frac{1}{9} \left(  \Xi_1 + \Xi_2  \right)$ \\ 
\hline
$K^{0}, \overline{K}^{0}$ & $\frac{g^2}{8} \left(  v_{\chi}^2  + v_{\eta}^2  \right)$ \\ 
\hline
\end{tabular}
\caption{Physical gauge bosons and their masses.}
\label{tab:gaugebosons}
\end{table}
\end{minipage}
\begin{minipage}{.5\textwidth}
$\Xi_1 = 3 g^2 \left(v _{\text{$\eta $}}^2+v
   _{\text{$\rho $}}^2+v _{\text{$\chi
   $}}^2\right)+\left(g'\right)^2 \left(v _{\text{$\eta
   $}}^2+4 v _{\text{$\rho $}}^2+v _{\text{$\chi
   $}}^2\right)$ \\ 
$\Xi_2=\sqrt{\begin{aligned}\left(3 g^2 \left(v _{\text{$\eta
   $}}^2+v _{\text{$\rho $}}^2+v _{\text{$\chi
   $}}^2\right)+\left(g'\right)^2 \left(v _{\text{$\eta
   $}}^2+4 v _{\text{$\rho $}}^2+v _{\text{$\chi
   $}}^2\right)\right){}^2 \\ -9 g^2 \left(3 g^2+4
   \left(g'\right)^2\right) \left(v _{\text{$\eta $}}^2
   \left(v _{\text{$\rho $}}^2+v _{\text{$\chi
   $}}^2\right)+v _{\text{$\rho $}}^2 v _{\text{$\chi
   $}}^2\right)\end{aligned}}$
\end{minipage}

with $v_{\eta} \simeq 173.948$ GeV, $v_{\rho} \simeq 173.948$ GeV and $v_{\chi} \simeq 10$ TeV. Consequently, for these values we find that the heavy gauge boson masses are $M_{W'} \approx 3.2$ TeV, $M_{Z'} \approx 6.3$ TeV. Notice that the obtained value of $M_{Z'} \approx 6.3$ TeV is consistent with the lower bound of $4$ TeV on the $Z^\prime $ gauge boson mass imposed by the  experimental data on the $K$, $D$ and $B$ meson mixings \cite{Huyen:2012uk}. 

In what follows we briefly comment about the LHC signals of a $Z^{\prime}$ gauge boson. The heavy $Z^{\prime}$ gauge boson is mainly produced via Drell-Yan mechanism and its corresponding production cross section has been found to range from $85$ fb up to $10$ fb for $Z^\prime $ gauge boson masses between $4$ TeV and $5$ TeV and LHC center of mass energy $\sqrt{S}=13$ TeV \cite{Long:2018dun}. Such $Z^\prime$ gauge boson after being produced will decay into pair of SM particles, with dominant decay mode into quark-antiquark pairs as shown in detail in Refs. \cite{Perez:2004jc,CarcamoHernandez:2005ka}. Comprehensive studies of the two body decays of the $Z^\prime$ gauge boson in 3-3-1 models are performed in Refs. \cite{Perez:2004jc,CarcamoHernandez:2005ka}, where it has been shown that the branching ratios of the $Z^\prime$ decays into a lepton pair are of the order of $10^{-2}$, thus yielding a total LHC cross section of about $1$ fb for the $pp\to Z^\prime\to l^{+}l^{-}$ resonant production at $\protect\sqrt{S}=13$ TeV and $M_{Z^\prime}=4$ TeV, which is below its corresponding lower experimental bound resulting from LHC searches \cite{Aaboud:2017sjh}. Finally, as pointed out in Ref. \cite{Long:2018dun}, at the proposed energy upgrade of the LHC with $\protect\sqrt{S}=28$ TeV, the LHC production cross section for the $pp\to Z^\prime\to l^{+}l^{-}$ resonant production will be of the order of $10^{-2}$, at $M_{Z^\prime}=4$ TeV, which falls in the order of magnitude of its corresponding experimental lower bound resulting from LHC searches.

\subsection{The low energy scalar potential and scalar mass spectrum}
\label{scalarpotential} 
In our model, the renormalizable low energy scalar potential reads:
\begin{eqnarray}
V &=&-\mu _{\chi }^{2}(\chi ^{\dagger }\chi )-\mu _{\eta }^{2}(\eta
^{\dagger }\eta )-\mu _{\rho }^{2}(\rho ^{\dagger }\rho )+f\left( \eta
_{i}\chi _{j}\rho _{k}\varepsilon ^{ijk}+H.c.\right) +\lambda _{1}(\chi
^{\dagger }\chi )(\chi ^{\dagger }\chi )+\lambda _{2}(\rho ^{\dagger }\rho
)(\rho ^{\dagger }\rho )+\lambda _{3}(\eta ^{\dagger }\eta )(\eta ^{\dagger
}\eta )  \notag \\
&&+\lambda _{4}(\chi ^{\dagger }\chi )(\rho ^{\dagger }\rho )+\lambda
_{5}(\chi ^{\dagger }\chi )(\eta ^{\dagger }\eta )+\lambda _{6}(\rho
^{\dagger }\rho )(\eta ^{\dagger }\eta )+\lambda _{7}(\chi ^{\dagger }\eta
)(\eta ^{\dagger }\chi )+\lambda _{8}(\chi ^{\dagger }\rho )(\rho ^{\dagger
}\chi )+\lambda _{9}(\rho ^{\dagger }\eta )(\eta ^{\dagger }\rho ),  \label{V}
\end{eqnarray}
being $\chi$, $\rho$ and $\eta$, the $SU(3)_L$ scalar triplets.
 
The following relations arise from the global minimal conditions of the low energy scalar potential:
\begin{eqnarray}
\mu_\chi^2 &=& -\frac{f v _{\text{$\eta $}} v _{\text{$\rho$}}}{\sqrt{%
2} v _{\text{$\chi $}}}+\frac{1}{2} \lambda _5 v _{\text{$\eta $}}^2+%
\frac{1}{2} \lambda _4 v _{\text{$\rho $}}^2+\lambda _1 v _{\text{$\chi 
$}}^2,\\
\mu_\rho^2 &=& -\frac{f v _{\text{$\eta $}} v _{\text{$\chi $}}}{\sqrt{%
2} v _{\text{$\rho $}}}+\frac{1}{2} \lambda _6 v _{\text{$\eta $}%
}^2+\lambda _2 v _{\text{$\rho $2}}^2+\frac{1}{2} \lambda _4 v _{\text{$%
\chi $}}^2, \\
\mu_\eta^2 &=& -\frac{f v _{\text{$\rho $}} v _{\text{$\chi $}}}{\sqrt{%
2} v _{\text{$\eta $}}}+\lambda _3 v _{\text{$\eta $}}^2+\frac{1}{2}
\lambda _6 v _{\text{$\rho $}}^2+\frac{1}{2} \lambda _5 v _{\text{$\chi 
$}}^2,
\end{eqnarray}
From the scalar potential we find that the squared scalar mass matrices are:
\begin{align*}
M^2_{\zeta\zeta} = 
\begin{pmatrix}
 \frac{f v _{\text{$\eta $}}
   v _{\text{$\rho
   $}}}{\sqrt{2} v
   _{\text{$\chi $}}} & \frac{f
   v _{\text{$\eta
   $}}}{\sqrt{2}} & \frac{f v
   _{\text{$\rho $}}}{\sqrt{2}}
   \\
 \frac{f v _{\text{$\eta
   $}}}{\sqrt{2}} & \frac{f v
   _{\text{$\eta $}} v
   _{\text{$\chi $}}}{\sqrt{2}
   v _{\text{$\rho $}}} &
   \frac{f v _{\text{$\chi
   $}}}{\sqrt{2}} \\
 \frac{f v _{\text{$\rho
   $}}}{\sqrt{2}} & \frac{f v
   _{\text{$\chi $}}}{\sqrt{2}}
   & \frac{f v _{\text{$\rho
   $}} v _{\text{$\chi
   $}}}{\sqrt{2} v
   _{\text{$\eta $}}}\\
  \end{pmatrix},  
   & & M^2_{\xi\xi} = \begin{pmatrix}
 2 \lambda _1 v _{\text{$\chi
   $}}^2+\frac{f v
   _{\text{$\eta $}} v
   _{\text{$\rho $}}}{\sqrt{2}
   v _{\text{$\chi $}}} &
   \lambda _4 v _{\text{$\rho
   $}} v _{\text{$\chi
   $}}-\frac{f v _{\text{$\eta
   $}}}{\sqrt{2}} & \lambda _5
   v _{\text{$\eta $}} v
   _{\text{$\chi $}}-\frac{f v
   _{\text{$\rho $}}}{\sqrt{2}}
   \\
 \lambda _4 v _{\text{$\rho
   $}} v _{\text{$\chi
   $}}-\frac{f v _{\text{$\eta
   $}}}{\sqrt{2}} & 2 \lambda _2
   v _{\text{$\rho
   $}}^2+\frac{f v
   _{\text{$\eta $}} v
   _{\text{$\chi $}}}{\sqrt{2}
   v _{\text{$\rho $}}} &
   \lambda _6 v _{\text{$\eta
   $}} v_{\text{$\rho
   $}}-\frac{f v _{\text{$\chi
   $}}}{\sqrt{2}} \\
 \lambda _5 v _{\text{$\eta
   $}} v _{\text{$\chi
   $}}-\frac{f v _{\text{$\rho
   $}}}{\sqrt{2}} & \lambda _6
   v _{\text{$\eta $}} v
   _{\text{$\rho $}}-\frac{f v
   _{\text{$\chi $}}}{\sqrt{2}}
   & 2 \lambda _3 v
   _{\text{$\eta $}}^2+\frac{f
   v _{\text{$\rho $}} v
   _{\text{$\chi $}}}{\sqrt{2}
   v _{\text{$\eta $}}}, \\
\end{pmatrix},
\end{align*}
\begin{align*}
M^2_{\chi^0_1\eta^0_3} = M^2_{\bar{\chi}^0_1\bar{\eta}^0_3} = \begin{pmatrix}
 \lambda _7 v _{\text{$\eta
   $}}^2+\frac{\sqrt{2} f v
   _{\text{$\rho $}} v
   _{\text{$\eta $}}}{v
   _{\text{$\chi $}}} & \sqrt{2}
   f v _{\text{$\rho
   $}}+\lambda _7 v
   _{\text{$\eta $}} v
   _{\text{$\chi $}} \\
 \sqrt{2} f v _{\text{$\rho
   $}}+\lambda _7 v
   _{\text{$\eta $}} v
   _{\text{$\chi $}} & \lambda
   _7 v _{\text{$\chi
   $}}^2+\frac{\sqrt{2} f v
   _{\text{$\rho $}} v
   _{\text{$\chi $}}}{v
   _{\text{$\eta $}}} \\
\end{pmatrix}, & & M^2_{\eta^{\pm}_2\rho^{\pm}_1} =
\begin{pmatrix}
 \lambda _9 v _{\text{$\rho
   $}}^2+\frac{\sqrt{2} f v
   _{\text{$\chi $}} v
   _{\text{$\rho $}}}{v
   _{\text{$\eta $}}} & \lambda
   _9 v _{\text{$\eta $}} v
   _{\text{$\rho $}}+\sqrt{2} f
   v _{\text{$\chi $}} \\
 \lambda _9 v _{\text{$\eta
   $}} v _{\text{$\rho
   $}}+\sqrt{2} f v
   _{\text{$\chi $}} & \lambda
   _9 v _{\text{$\eta
   $}}^2+\frac{\sqrt{2} f v
   _{\text{$\chi $}} v
   _{\text{$\eta $}}}{v
   _{\text{$\rho $}}} \\
\end{pmatrix},
\end{align*}
\begin{align}
    M^{2}_{\chi^{\pm}_2\rho^{\pm}_3} = \begin{pmatrix}
   \lambda _8 v _{\text{$\rho
   $}}^2+\frac{\sqrt{2} f v
   _{\text{$\eta $}} v
   _{\text{$\rho $}}}{v
   _{\text{$\chi $}}} & \sqrt{2}
   f v _{\text{$\eta
   $}}+\lambda _8 v
   _{\text{$\rho $}} v
   _{\text{$\chi $}} \\
 \sqrt{2} f v _{\text{$\eta
   $}}+\lambda _8 v
   _{\text{$\rho $}} v
   _{\text{$\chi $}} & \lambda
   _8 v _{\text{$\chi
   $}}^2+\frac{\sqrt{2} f v
   _{\text{$\eta $}} v
   _{\text{$\chi $}}}{v
   _{\text{$\rho $}}} \\
\end{pmatrix}.
\end{align}

The resulting physical scalars and their masses are given in Table \ref{tab:escalares}.

\begin{minipage}{0.55\textwidth}
\begin{table}[H]
\centering
\renewcommand{\arraystretch}{1.3}
\begin{tabular}{c|c}
\textbf{Scalars} & \textbf{Masses}  \\
\hline\hline\
$G_1^0 = -S_{\alpha} \zeta_{\chi} + C_{\alpha} \zeta_{\eta}$ & $M^2_{G_1^0}=0$ \\
\hline
$A^{0} = C_{\beta} \zeta_{\rho} + S_{\beta} \zeta_{\eta}$ & $M_{A^{0}} = \frac{f}{\sqrt{2}} v_{\chi}\left(\frac{v_{\rho}}{v_{\eta}}+\frac{v_{\eta}}{v_{\rho}}\right)$ \\
\hline
$G_2^0 = -C_{\gamma} \zeta_{\chi} + S_{\gamma}\zeta_{\rho}$ & $M^2_{G_2^0}=0$  \\
\hline
$H^{0}_{1} = \xi_{\chi} $  &  $M^2_{H^{0}_1} = \lambda_{1} v^{2}_{\chi}$  \\ 
\hline
$h^0 = C_{\delta}\xi_{\rho} -S_{\delta}\xi_{\eta} $  & $M^2_{h^0} = \lambda_{3} v^{2}_{\eta} + \lambda_{2} v^{2}_{\rho}$ \\ 
\hline
$H^{0}_{2} = S_{\delta} \xi_{\rho} + C_{\delta}\xi_{\eta}$  &  $M^2_{H^{0}_{2}} = \frac{fv_{\chi}}{\sqrt{2}}\left(\frac{v^{2}_{\eta}+v^{2}_{\rho}}{v_{\eta}v_{\rho}}\right)$  \\ 
\hline
$G^{0}_{3} = -C_{\alpha} \chi^{0}_1 + S_{\alpha} \eta^{0}_3$  & $M^2_{G^{0}_{3}} = 0$ \\ 
\hline
$\bar{G}_3^0 = -C_{\alpha} \bar{\chi}^{0}_1 + S_{\alpha} \bar{\eta}^{0}_3$  & $M^2_{\bar{G}^{0}_3} = 0$ \\ 
\hline
$H^{0}_{3} = S_{\alpha} \chi^{0}_1 + C_{\alpha} \eta^{0}_3$  &  $M^2_{H^{0}_{3}} =  \frac{(\sqrt{2}fv_{\rho}+\lambda_{7}v_{\eta}v_{\chi})(v^{2}_{\eta}+v^{2}_{\chi})}{v_{\eta}v_{\chi}}$ \\ 
\hline
$\bar{H}^{0}_{3} = S_{\alpha} \bar{\chi}^{0}_1 + C_{\alpha} \bar{\eta}^{0}_3$  &  $M^2_{\bar{H}^{0}_{3}} = M^2_{H^{0}_{3}} $ \\ 
\hline
$G^{\pm}_4 = -C_{\gamma} \chi^{\pm}_2 + S_{\gamma} \rho^{\pm}_3$  & $M^2_{G^{\pm}_4} = 0$ \\ 
\hline
$H^{\pm}_4 = S_{\gamma} \chi^{\pm}_2 + C_{\gamma} \rho^{\pm}_3$  & $M^2_{H^{\pm}_4} = \frac{(\sqrt{2}fv_{\eta} + \lambda_{8}v_{\rho}v_{\chi})(v^{2}_{\rho}+v^{2}_{\chi})}{v_{\rho}v_{\chi}}$ \\
\hline
$G^{\pm}_5 = -C_{\beta} \eta^{\pm}_2 + S_{\beta} \rho^{\pm}_1$  & $M^2_{G^{\pm}_5} = 0 $ \\ 
\hline
$H^{\pm}_5 = S_{\beta} \eta^{\pm}_2 + C_{\beta} \rho^{\pm}_1$  & $M^2_{H^{\pm}_5} = \frac{(\sqrt{2}fv_{\chi}+\lambda_{9}v_{\eta}v_{\rho})(v^{2}_{\eta}+v^{2}_{\rho})}{v_{\eta}v_{\rho}}$ \\ 
\hline
\end{tabular}
\caption{Physical scalar fields with their masses.}
\label{tab:escalares}
\end{table}
\end{minipage}
\begin{minipage}{4\textwidth}
$\tan(\alpha)=\dfrac{v_\eta}{v_\chi}$ \\ 
$\tan(\beta)=\dfrac{v_\rho}{v_\eta}$ \\
$\tan(\gamma)=\dfrac{v_\rho}{v_\chi}$ \\
$\tan(\delta)= \dfrac{2 \left(v_{\eta} \lambda_6 v_{\rho} -\frac{f v_{\chi}}{\sqrt{2}}\right)}{\frac{f v_{\eta}  v_{\chi}}{\sqrt{2} v_{\rho} }-\frac{f v_{\rho} v_{\chi} }{\sqrt{2} v_{\eta} }-2 v_{\eta}^2 \lambda_3 +2 \lambda_2 v_{\rho}^2}$
\end{minipage}

The field composition of the low energy physical scalar spectrum of our model is given by 
one light neutral scalar $h^0$ identified with the SM-like $125.09$ GeV Higgs boson found at the LHC, five neutral heavy Higgs bosons $(H_1^0,H_2^0,H_3^0,\bar{H}_3^0,A^0)$ and four charged Higgs bosons $(H^{\pm}_{4},H^{\pm}_{5})$. It's worth mentioning that the neutral
Goldstone bosons $G_1^0$, $G_2^0$, $G_3^0$ and $\overline{G}_3^0$ are related to the longitudinal components of the $Z$, $Z^{\prime }$, $K^0$
and $\overline{K}^0$ just like the charged Goldstone bosons $G^{\pm}_1$
and $G^{\pm}_2$ are related to the longitudinal components of the $W^\pm$
and $W'^\pm$ gauge bosons.

The $125$ GeV mass value for the SM-like Higgs boson can be reproduced for the following benchmark point:
\begin{eqnarray}
\begin{gathered}
v_{\chi} \simeq 10 \text{TeV}, \hspace{0.6cm} v_{\eta} \sim v_{\rho} \simeq 174 \text{GeV}, \hspace{0.6cm} f \simeq 1000 \text{ GeV},\\ 
\lambda_1 \simeq 0.016, \hspace{0.6cm} \lambda_2 \sim \lambda_3 \simeq 0.26, \hspace{0.6cm} \lambda_4 \sim \lambda_5 \sim \lambda_7 \simeq 1,  \hspace{0.6cm} \lambda_6 \simeq 10, \hspace{0.6cm} \lambda_8 \simeq 0.01.
\label{benchmark}
\end{gathered}
\end{eqnarray}
where in this scenario, the trilinear parameter $f$ has to be fixed at $1000$ GeV to get $M_{A^0} \simeq 5318.3$ GeV and $M_{H^{\pm}} \simeq 5503.95$ GeV. 

\section{Quark masses and mixings.}

\label{quarksector} From the Yukawa interactions of the quark sector given by Eq. (\ref%
{Lyq}), we find that the SM quark mass matrices read: 
\begin{equation}
M_{U}=\frac{v}{\sqrt{2}}\left( 
\begin{array}{ccc}
c_{1}\lambda ^{8} & 0 & a_{1}\lambda ^{4} \\ 
0 & b_{2}\lambda ^{4} & 0 \\ 
0 & 0 & a_{2}%
\end{array}%
\right) ,\hspace{1cm}\hspace{1cm}M_{D}=\frac{v}{\sqrt{2}}\left( 
\begin{array}{ccc}
g_{1}\lambda ^{7} & g_{4}\lambda ^{6} & 0 \\ 
0 & g_{2}\lambda ^{5} & 2g_{2}\lambda ^{5} \\ 
0 & 0 & g_{3}\lambda ^{3}%
\end{array}%
\right) ,  \label{Mq}
\end{equation}%
where $\lambda =0.225$ and $v=246$ GeV. In order to get quark mixing angles and a CP violating phase consistent with the experimental data, we assume that all dimensionless parameters given in Eqs. \eqref{Mq} are real, except for $a_1$, taken to be complex.

The exotic quark masses are: 
\begin{equation}
m_{t^{\prime}}=y^{\left(t^{\prime}\right) }\frac{v_{\chi }}{\sqrt{2}},\hspace{1cm}%
m_{j_{1}}=y_{1}^{\left(j\right) }\frac{v_{\chi }}{\sqrt{2}}=\frac{%
y_{1}^{\left( j\right) }}{y^{\left( t^{\prime}\right) }}m_{t^{\prime}},\hspace{1cm}%
m_{j_{2}}=y_{2}^{\left(j\right) }\frac{v_{\chi }}{\sqrt{2}}=\frac{%
y_{2}^{\left(j\right) }}{y^{\left( t^{\prime}\right) }}m_{t^{\prime}}.  \label{mexotics}
\end{equation}%

The experimental values of the physical quark mass spectrum \cite{Bora:2012tx,Xing:2007fb}, mixing angles and Jarlskog invariant \cite{Olive:2016xmw} are consistent with their experimental data, as shown in Table \ref{Tab:quarks}, starting from the following benchmark point:
\begin{eqnarray}
\begin{gathered}
c_{1} \simeq 1.2525,\hspace{0.6cm}\left\vert a_{1}\right\vert \simeq 1.48406, \hspace{0.6cm}\arg \left( a_{1}\right) \simeq 68^{\circ },\hspace{0.6cm} a_{2}\simeq 0.989375,\hspace{0.6cm}b_{2}\simeq 1.41504,  \\
g_{1} \simeq 0.579397,\hspace{1cm} g_{2}\simeq 0.57,%
\hspace{1cm} g_{3}\simeq 1.40209,\hspace{1cm}%
g_{4}\simeq 0.583.  \label{eq:Quark-benchmark-point}
\end{gathered}
\end{eqnarray}
The result given in Eq. \eqref{eq:Quark-benchmark-point} motivates to consider the simplified benchmark scenario: 
\begin{eqnarray}
\begin{gathered}
c_{1} \simeq 1.2525,\hspace{0.6cm}\left\vert a_{1}\right\vert \simeq 1.48406, \hspace{0.6cm}\arg \left( a_{1}\right) \simeq 68^{\circ },\hspace{0.6cm} a_{2}\simeq 0.989375,\hspace{0.6cm} \\
g_{1} \simeq g_{2}\simeq g_{4} \simeq 0.579397,
\hspace{1cm} b_{2} \simeq g_{3} \simeq 1.41504.
  \label{eq:Quark-benchmark-point2}
\end{gathered}
\end{eqnarray}
Notice that a successful fit of the ten physical observables in the quark sector can be obtained in the above described scenarios where the first (Eq. (\ref{eq:Quark-benchmark-point})) and the second one (Eq. (\ref{eq:Quark-benchmark-point2})) only have 9 and 6 effective free parameters, respectively. It is worth mentioning that in the general scenario of 9 effective free parameters, such parameters fit the CKM quark mixing matrix as well as 5 of the 6 quark masses, whereas the remaining quark mass is predicted. In the concerning to the scenario of 6 effective free parameters, the quark mixing angle $\theta_{13}$, the quark CP violating phase $\delta$ and 2 quark masses are adjusted, whereas the remaining quark mixing angles $\theta_{12}$, $\theta_{23}$ and 4 quark masses are predicted.

Thus, the symmetries of our model give rise to quark mass matrix textures that successfully explain the SM quark mass spectrum and mixing parameters, with quark sector effective free parameters of order unity.

\begin{table}[H]
\begin{center}
\begin{tabular}{c|l|l|l}
\hline\hline
Observable & Model value with Eq. \eqref{eq:Quark-benchmark-point} & Model value with Eq. \eqref{eq:Quark-benchmark-point2} & Experimental value \\ \hline
$m_{u}(MeV)$ & \quad $1.44999$ & \quad $1.44999$ & \quad $1.45_{-0.45}^{+0.56}$ \\ \hline
$m_{c}(MeV)$ & \quad $635$ & \quad $635$ & \quad $635\pm 86$ \\ \hline
$m_{t}(GeV)$ & \quad $172.101$ & \quad $172.101$ & \quad $172.1\pm 0.6\pm 0.9$ \\ \hline
$m_{d}(MeV)$ & \quad $2.89988$ & \quad $2.90313$ & \quad $2.9_{-0.4}^{+0.5}$ \\ \hline
$m_{s}(MeV)$ & \quad $59.1145$ & \quad $60.021$ & \quad $57.7_{-15.7}^{+16.8}$ \\ \hline
$m_{b}(GeV)$ & \quad $2.79418$ & \quad $2.82003$ & \quad $2.82_{-0.04}^{+0.09}$ \\ \hline
$\sin \theta_{12}$ & \quad $0.225402$ & \quad $0.220611$ & \quad $0.225$ \\ \hline
$\sin \theta_{23}$ & \quad $0.0412799$ & \quad $0.0415761$ & \quad $0.0412$ \\ \hline
$\sin \theta_{13}$ & \quad $0.00386484$ & \quad $0.0038648$ & \quad $0.00351$ \\ \hline
$\delta_{q}$ & \quad $68.021^{\circ}$ & \quad $68.0198^{\circ}$ & \quad $68^{\circ}$ \\ \hline\hline
\end{tabular}%
\end{center}
\caption{Model and experimental values of the quark masses and CKM parameters.}
\label{Tab:quarks}
\end{table}

\begin{figure}[H]
	\begin{subfigure}{.5\linewidth}
	\centering
    \captionsetup{width=0.8\textwidth}
	\includegraphics[scale=0.55]{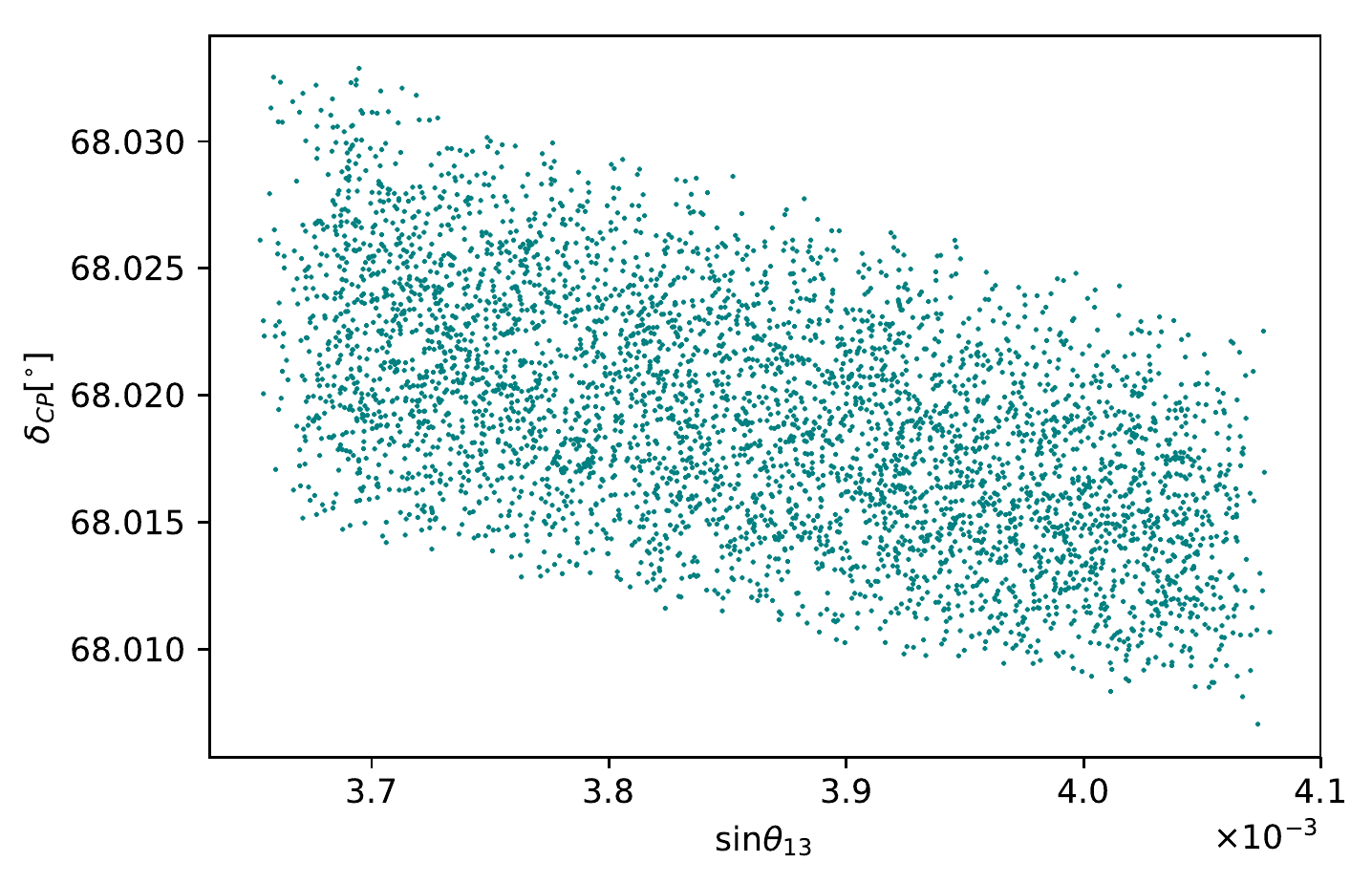}
	\caption{Correlation between $\delta_{CP}$ and $\sin\theta_{13}$.}
	\label{fig:subquark1}
	\end{subfigure}
	\begin{subfigure}{.5\linewidth}
	\centering
    \captionsetup{width=0.8\textwidth}
	\includegraphics[scale=0.55]{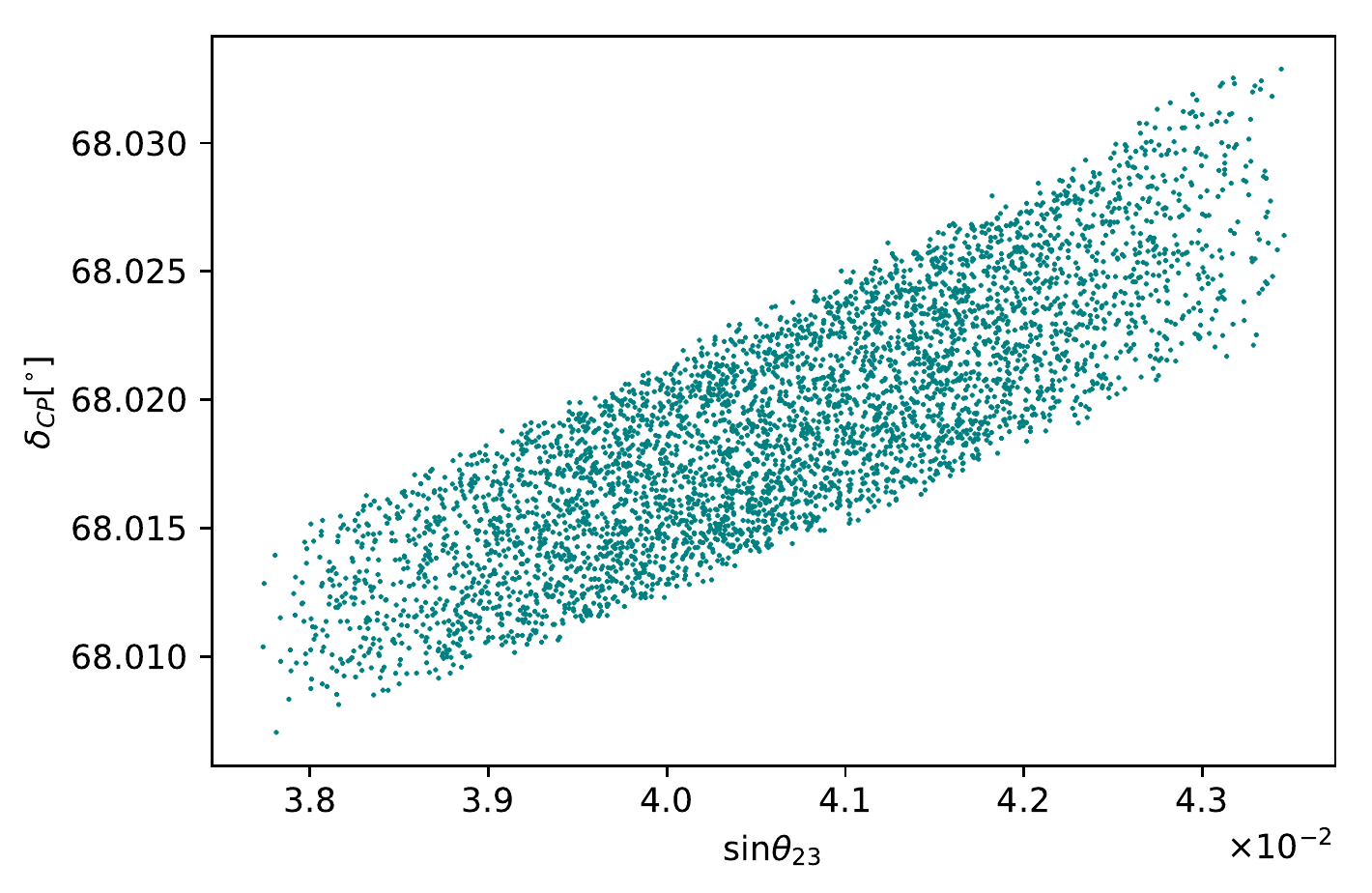}
	\caption{Correlation between $\delta_{CP}$ and $\sin\theta_{23}$.}
	\label{fig:subquark2}
	\end{subfigure}
	\caption{Correlations between the quark CP-violating phase $\delta_{CP}$ and the quark mixing parameters $\sin \theta_{13}$, $\sin \theta_{23}$.}
	\label{fig:quarkcorrelation}
\end{figure}

In addition to the benchmark points of Eqs. \eqref{eq:Quark-benchmark-point} and (\ref{eq:Quark-benchmark-point2}), correlation plots in Figure \ref{fig:quarkcorrelation} have been obtained to analyze the behaviour of some of the quark observables, such as the CP violating phase $\delta_{CP}$ as a function of the quark mixing parameters $\sin\theta_{13}$ and $\sin\theta_{23}$.
These plots were generated by varying the quark sector parameters in Eq. \eqref{eq:Quark-benchmark-point} in a range of values that satisfies the $3\sigma$ experimental allowed values in the quark sector and $0.224<\sin\theta_{12}< 0.226$. The plots show that the CP-violating phase is predicted to be in range $68.007^{\circ} \lesssim \delta_{CP} \lesssim 68.032^{\circ}$ for the allowed parameter space. Figure \ref{fig:subquark1} shows that $0.0036 \lesssim \sin\theta_{13} \lesssim 0.0040 $ and as $\sin\theta_{13}$ grows up $\delta_{CP}$ goes down. On the other hand, Figure \ref{fig:subquark2} shows that $0.0377 \lesssim \sin\theta_{23} \lesssim 0.0434 $ and $\delta_{CP}$ is directly proportional to $\sin\theta_{23}$.

Finally to close this section we briefly comment about the LHC signatures of exotic $t^{\prime}$, $j_{1}$ and $j_{2}$ quarks in our model. Such exotic quarks will mainly decay into a top quark and either
neutral or charged scalar and can be pair produced at the LHC via Drell-Yan and gluon fusion processes mediated by charged gauge bosons and gluons, respectively. A detailed study of the collider phenomenology of the model is beyond the scope of this paper and is left for future studies.

\section{Lepton masses and mixings.}

\label{leptonsector}

From Eq. (\ref{Lyl}), and using the product rules of the $S_4$ group given in Appendix \ref{appen:s4}, we find that the charged lepton mass matrix is given by: 
\begin{equation}
M_{l}=\frac{v}{\sqrt{2}}\left( 
\begin{array}{ccc}
x_{1}\lambda ^{9} & x_{4}\lambda ^{5} & x_{5}\lambda ^{3} \\ 
0 & x_{2}\lambda ^{5} & x_{6}\lambda^{4} \\ 
0 & 0 & x_{3}\lambda ^{3}%
\end{array}%
\right) .
\end{equation}%
Regarding the neutrino sector, from Eq. (\ref{Lyl}), we find the following neutrino mass terms: 
\begin{equation}
-\mathcal{L}_{mass}^{\left( \nu \right) }=\frac{1}{2}\left( 
\begin{array}{ccc}
\overline{\nu _{L}^{C}} & \overline{\nu _{R}} & \overline{N_{R}}%
\end{array}%
\right) M_{\nu }\left( 
\begin{array}{c}
\nu _{L} \\ 
\nu _{R}^{C} \\ 
N_{R}^{C}%
\end{array}%
\right) +H.c,  \label{Lnu}
\end{equation}%
where the neutrino mass matrix is given by: 
\begin{equation}
M_{\nu }=\left( 
\begin{array}{ccc}
0_{3\times 3} & M_{1} & M_{2} \\ 
M_{1}^{T} & 0_{3\times 3} & M_{3} \\ 
M_{2}^{T} & M_{3}^{T} & 0_{3\times 3}%
\end{array}%
\right) ,  \label{Mnu}
\end{equation}%
and the submatrices take the form: 
\begin{eqnarray}
M_{1} &=&\frac{h_{\rho }v_{\rho }v_{\zeta }}{2\Lambda }\left( 
\begin{array}{ccc}
0 & a & 0 \\ 
-a & 0 & b \\ 
0 & -b & 0%
\end{array}%
\right) ,\hspace{0.7cm}\hspace{0.7cm}M_{2}=h_{\eta }^{\left( L\right) }\frac{%
v_{\eta }v_{\Sigma }}{\sqrt{6}\Lambda }\left( 
\begin{array}{ccc}
x & y & -y \\ 
-x & \omega ^{2}y & -\omega y \\ 
x & \omega y & -\omega ^{2}y%
\end{array}%
\right) ,  \notag \\
M_{3} &=&h_{\chi }^{\left( L\right) }\frac{v_{\chi }v_{\Sigma }}{\sqrt{6}%
\Lambda }\left( 
\begin{array}{ccc}
r & z & -z \\ 
-r & \omega ^{2}z & -\omega z \\ 
r & \omega z & -\omega ^{2}z%
\end{array}%
\right) ,\hspace{0.7cm}\hspace{0.7cm}\omega =e^{\frac{2\pi i}{3}}.
\end{eqnarray}%

The light active neutrino masses arise from a linear seesaw mechanism and the physical neutrino mass matrices are: 
\begin{eqnarray}
M_{\nu }^{\left( 1\right) } &=&-\left[ M_{2}M_{3}^{-1}M_{1}^{T}+M_{1}\left(
M_{3}^{T}\right) ^{-1}M_{2}^{T}\right] , \\
M_{\nu }^{\left( 2\right) } &=&-\frac{1}{2}\left( M_{3}+M_{3}^{T}\right) -%
\frac{1}{2}\left[ M_{1}^{T}M_{1}\left( M_{3}^{T}\right) ^{-1}+\left(
M_{3}\right) ^{-1}M_{1}^{T}M_{1}\right] , \\
M_{\nu }^{\left( 3\right) } &=&\frac{1}{2}\left( M_{3}+M_{3}^{T}\right) +%
\frac{1}{2}\left[ M_{1}^{T}M_{1}\left( M_{3}^{T}\right) ^{-1}+\left(
M_{3}\right) ^{-1}M_{1}^{T}M_{1}\right] ,
\end{eqnarray}%
where $M_{\nu }^{\left( 1\right) }$ corresponds to the active neutrino mass
matrix whereas $M_{\nu }^{\left( 2\right) }$ and $M_{\nu }^{\left( 3\right)
} $ are the sterile neutrino mass matrices. The physical neutrino spectrum is composed of 3 light active neutrinos and 6 nearly degenerate sterile
exotic pseudo-Dirac neutrinos. Furthermore, from Eqs. (\ref{Lyl}) and (\ref{VEVsinglets}) and considering $v_{\chi }\sim\mathcal{O}(10)$ TeV, $v_{\eta}\sim v_{\rho}\sim {O}(100)$ GeV and the Yukawa couplings of order unity, we find that the light active neutrino mass scale $\sim 50$ meV is estimated as $m_{\nu}\sim \frac{v_{\eta}v_{\rho}v_{\zeta}}{v_{\chi}\Lambda}\sim \frac{v_{\eta}v_{\rho}}{\Lambda}$, which implies for the model cutoff the estimate $\Lambda\sim\mathcal{O}(10^{16})$ GeV.

The sterile neutrinos can be produced in pairs at the LHC, via quark-antiquark annihilation mediated by a heavy $Z^\prime $ gauge boson. They can decay into SM particles giving rise to a SM charged lepton and a $W$ gauge boson in the final state. Thus, observing an excess of events with respect to the SM background in the opposite sign dileptons final states can be a signal in support of this model at the LHC. Studies of inverse seesaw neutrino signatures at colliders as well as the production of heavy neutrinos at the LHC are carried out in Refs. \cite{Dev:2009aw,BhupalDev:2012zg,Das:2012ze,Dev:2013oxa,Das:2014jxa,Das:2016hof,Das:2017gke,Das:2017nvm,Das:2017zjc,Das:2017rsu,Das:2018usr,Das:2018hph,Bhardwaj:2018lma,Helo:2018rll,Pascoli:2018heg}. A detailed study of the implications of our model at colliders goes beyond the scope of this paper and is deferred for a future work.

The light active neutrino mass matrix is given by: 
\begin{equation}
M_{\nu }^{\left( 1\right) }=\left(
\begin{array}{ccc}
 2 A & B e^{i \varphi }-2 A & A-B e^{i \varphi } \\
 B e^{i \varphi }-2 A & 2 \left(A-B e^{i \varphi }\right) & 2 B e^{i \varphi }-A \\
 A-B e^{i \varphi } & 2 B e^{i \varphi }-A & -2 B e^{i \varphi } \\
\end{array}
\right).
\end{equation}%
and the light active neutrino masses are: 
\begin{eqnarray}
m_{1} &=&0, \\
m_{2} &=&\sqrt{2} \sqrt{5 A^2-2 \sqrt{6} \sqrt{A^4-3 A^3 B \cos (\varphi )+A^2 B^2 \cos (2 \varphi )+3 A^2 B^2-3 A B^3 \cos (\varphi )+B^4}-7 A B \cos (\varphi )+5 B^2},\notag\\
m_{3} &=&\sqrt{2} \sqrt{5 A^2+2 \sqrt{6} \sqrt{A^4-3 A^3 B \cos (\varphi )+A^2 B^2 \cos (2 \varphi )+3 A^2 B^2-3 A B^3 \cos (\varphi )+B^4}-7 A B \cos (\varphi )+5 B^2},\notag
\end{eqnarray}
which implies that the experimental values of the neutrino mass squared splittings can be very well reproduced for the following benchmark point:
\begin{equation}
A=B= 0.00949663\hspace{0.2mm}\textit{eV},\hspace{1cm}\varphi=65.8796^{\circ}. 
\end{equation}
The corresponding PMNS leptonic mixing matrix is defined as $U=R_{l}^{\dagger }R_{\nu }$, and from the standard parametrization of $U$, it follows that the lepton mixing parameters are given by:
\begin{equation*}
\sin ^{2}(\theta _{13})=|U_{13}|^{2},\hspace{1cm}\sin ^{2}(\theta _{12})=%
\dfrac{|U_{12}|^{2}}{1-|U_{13}|^{2}},\hspace{1cm}\sin ^{2}(\theta _{23})=%
\dfrac{|U_{23}|^{2}}{1-|U_{13}|^{2}}.
\end{equation*}
It is worth mentioning that due to the complexity of the expression for the PMNS matrix, the analytic form cannot be shown.

Furthermore, the Jarlskog invariant $J_{CP}$ is determined from the relation:
\begin{align}
J_{CP}=\text{Im}(U_{11}^{\ast }U_{23}^{\ast }U_{13}U_{21}),\label{eqn:jarlskog}
\end{align}
whereas the leptonic Dirac CP violating phase $\delta_{CP}$ can be extracted from the equivalent definition of $J_{CP}$ \cite{Krastev:1988yu} in the standard parametrization: 
\begin{align}
J_{CP}=\dfrac{1}{8} \sin (2\theta_{12})\sin (2\theta_{23})\sin (2\theta_{13})\cos (\theta_{13})\delta_{CP}.\label{eqn:deltacp}
\end{align}

The charged lepton masses, leptonic mixing parameters and CP-phase can be very well reproduced for the scenario of normal neutrino mass ordering in terms of natural parameters of order one, as shown in Table \ref{Tab:neutrinofit}, starting from the following benchmark point: 
\begin{eqnarray}
&&x_1=-0.85677-2.19346i,\hspace{1cm}x_2=-2.84582-1.22066i,\hspace{1cm}x_3=-0.235108-0.00451549i,\notag\\
&&x_4=0.979847+2.04567i,\hspace{1cm}x_5=0.220533+0.440278i,\hspace{1cm}x_6=-2.69931-1.24421i.
\end{eqnarray}
As indicated by Table \ref{Tab:neutrinofit}, our model is consistent with the experimental data on lepton masses and mixings. Notice that the ranges for the experimental values in Table \ref{Tab:neutrinofit} were taken from \cite{deSalas:2017kay} for the case of normal hierarchy. Note that we only consider the case of normal hierarchy since it is favored over more than $3\sigma$ than the inverted neutrino mass ordering.

\begin{figure}[H]
    \begin{subfigure}{.5\linewidth}
        \centering
        \captionsetup{width=0.8\textwidth}
        \includegraphics[scale=.4]{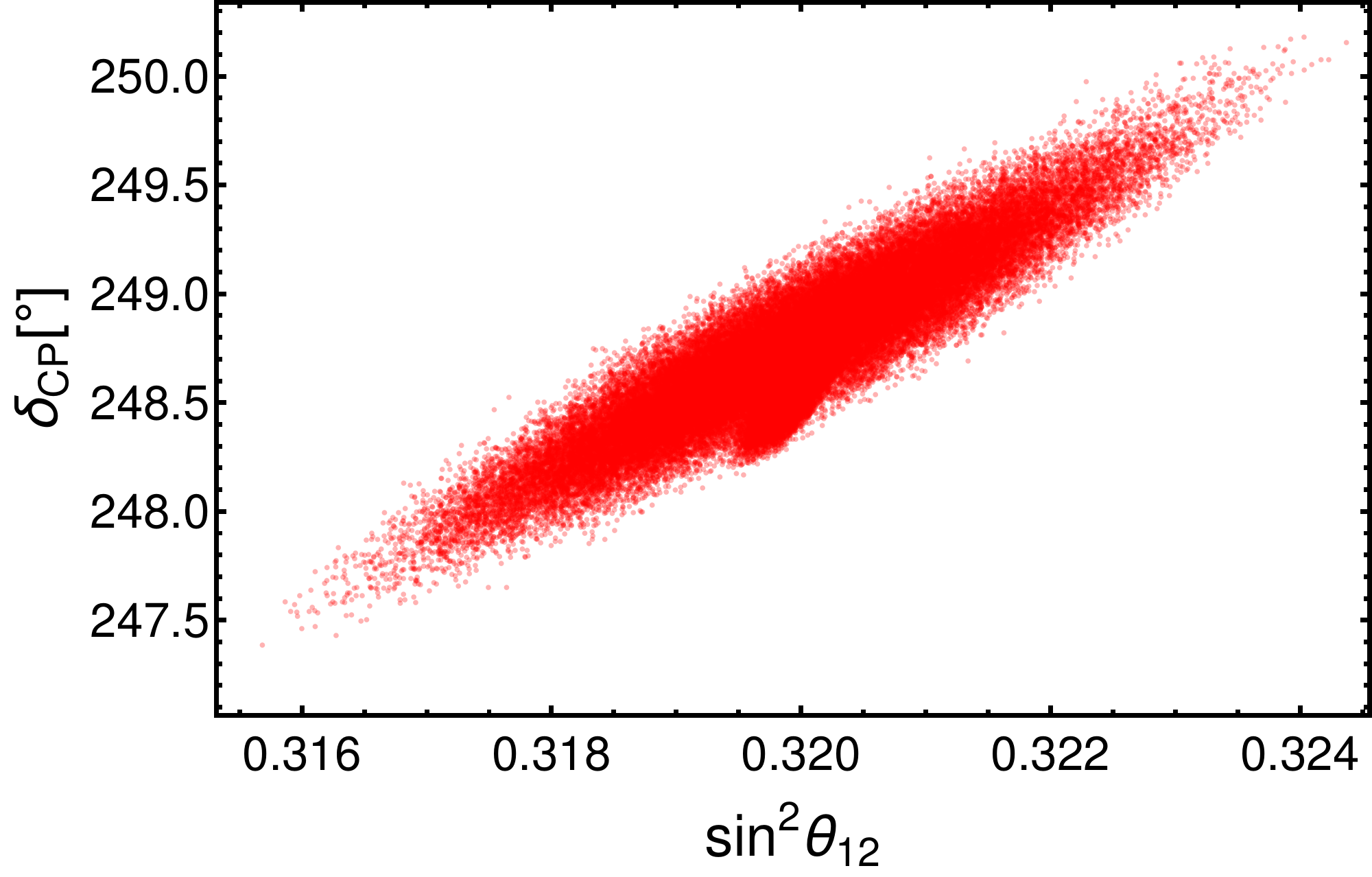}
        \caption{Correlation between the leptonic Dirac CP-violating phase $\delta_{CP}$ and the solar mixing parameter $\sin^2 \theta_{12}$.}
        \label{fig:sub1}
    \end{subfigure}
    \begin{subfigure}{.5\linewidth}
        \centering
        \captionsetup{width=0.8\textwidth}
        \includegraphics[scale=.4]{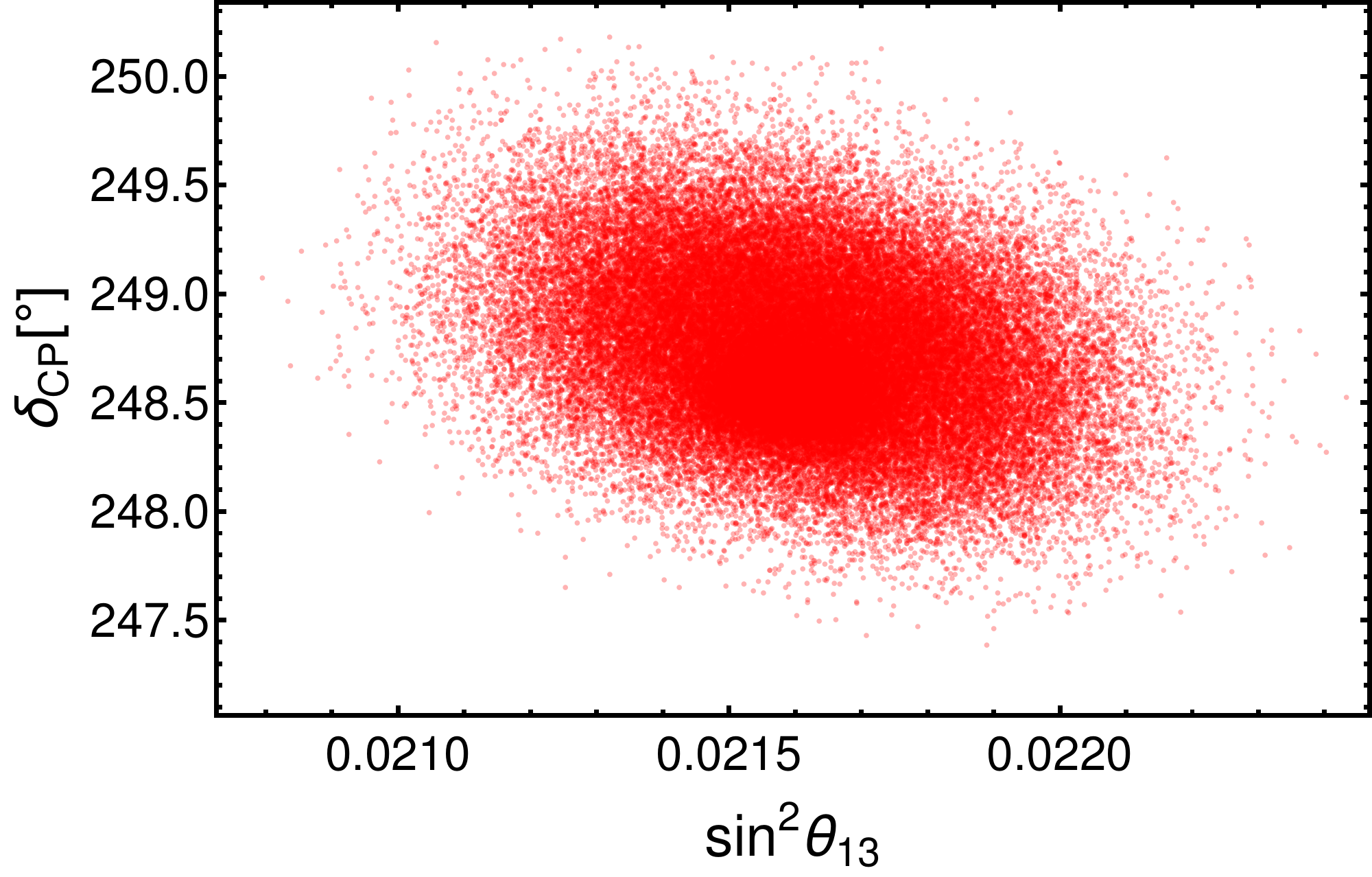}
        \caption{Correlation between the leptonic Dirac CP-violating phase $\delta_{CP}$ and the reactor mixing parameter $\sin^2 \theta_{13}$.}
        \label{fig:sub2}
    \end{subfigure}\\
    \begin{subfigure}{.5\linewidth}
        \centering
        \captionsetup{width=0.8\textwidth}
        \includegraphics[scale=.4]{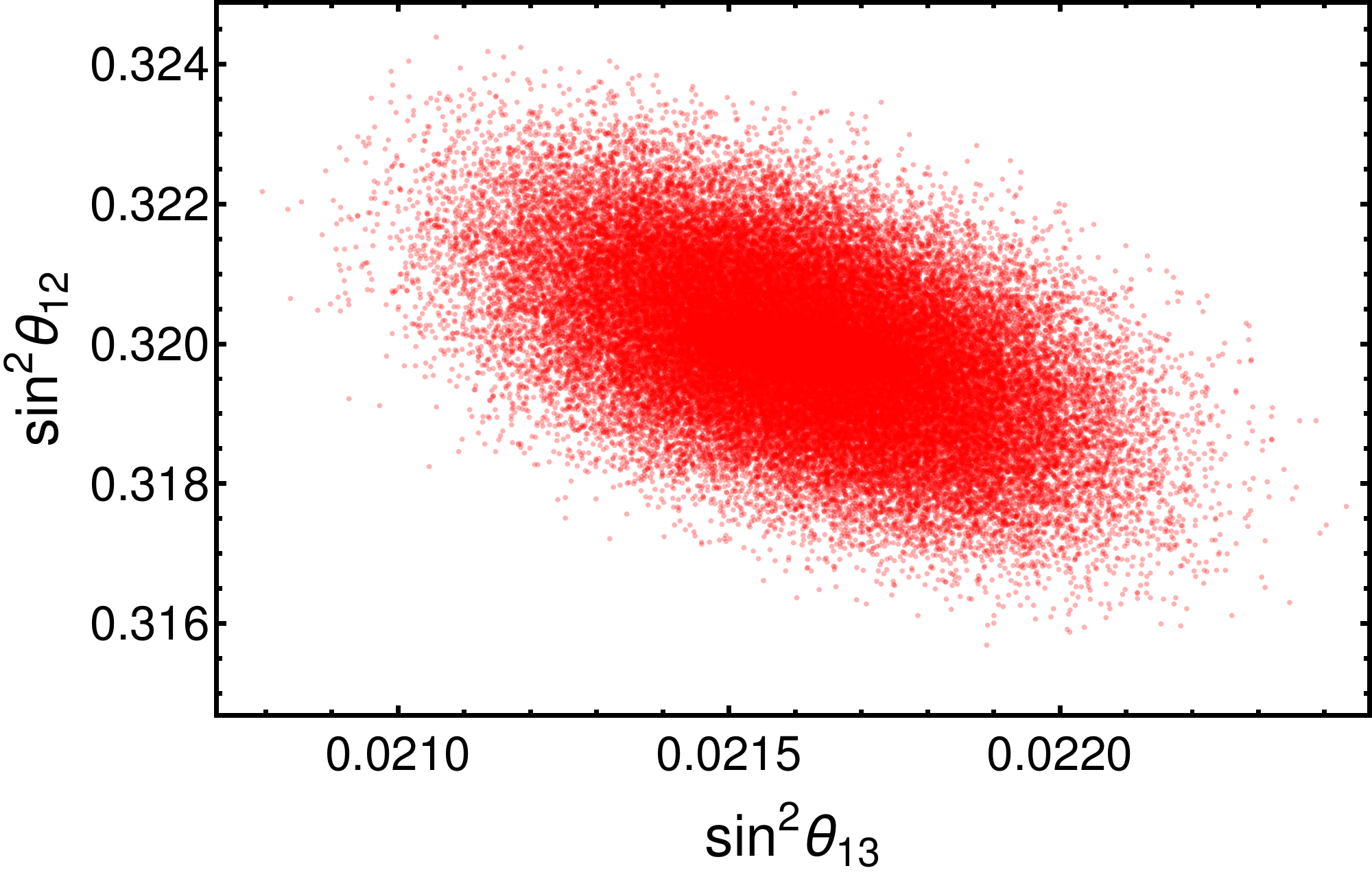}
        \caption{Correlation between the solar $\sin^2 \theta_{12}$ and reactor $\sin^2 \theta_{13}$ mixing parameters.}
        \label{fig:sub3}
    \end{subfigure}
    \begin{subfigure}{.5\linewidth}
        \centering
        \captionsetup{width=0.8\textwidth}
        \includegraphics[scale=.4]{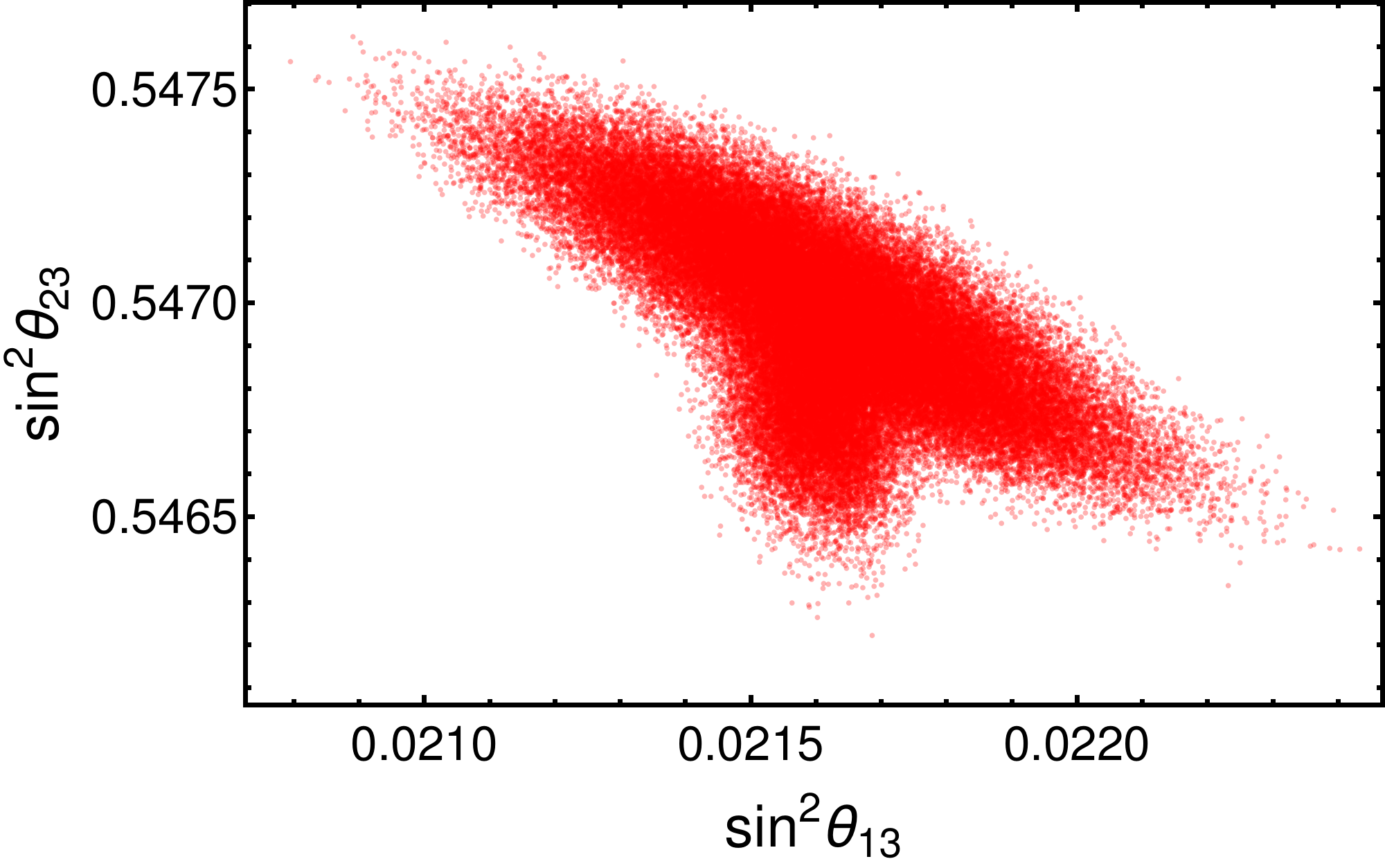}
        \caption{Correlation between the atmospheric $\sin^2 \theta_{23}$ and reactor $\sin^2 \theta_{13}$ mixing parameters.}
        \label{fig:sub3}
    \end{subfigure}
    \caption{Correlations between the different lepton sector observables.} 
    \label{fig:correlationleptonsector}
\end{figure}

\begin{table}[H]
\centering
\captionsetup{width=0.6\textwidth} 
\begin{tabular}{|c|c|c|c|c|}
\hline
\multirow{2}{*}{Observable} & \multirow{2}{*}{Model Value} & 
\multicolumn{3}{c|}{Experimental value} \\ \cline{3-5}
&  & $1\sigma$ range & $2\sigma$ range & $3\sigma$ range \\ \hline\hline
$m_e$ [MeV] & $0.487$ & $0.487$ & $0.487$ & $0.487$ \\ \hline
$m_\mu$ [MeV] & $102.8$ & $102.8 \pm 0.0003$ & $102.8 \pm 0.0006$ & $102.8 \pm 0.0009$ \\ \hline
$m_\tau$ [GeV] & $1.75$ & $1.75 \pm 0.0003$ & $1.75 \pm 0.0006$ & $1.75 \pm 0.0009$ \\ \hline
$\Delta m_{21}^2$ $[10^{-5}\,eV^2]$ & $7.55$ & $7.55^{+0.20}_{-0.16}$ & $7.20 -
7.94$ & $7.05 - 8.14$ \\ \hline
$\Delta m_{31}^2$ $[10^{-3}\,eV^2]$ & $2.50$ & $2.50\pm0.03$ & $2.44 - 2.57$ & $%
2.41-2.60$ \\ \hline
$\sin^2(\theta_{12})/10^{-1}$ & $3.20$ & $3.20^{+0.20}_{-0.16}$ & $%
2.89-3.59$ & $2.73-3.79$ \\ \hline
$\sin^2(\theta_{23})/10^{-1}$ & $5.47$ & $5.47^{+0.20}_{-0.30}$ & $%
4.67-5.83$ & $4.45-5.99$ \\ \hline
$\sin^2(\theta_{13})/10^{-2}$ & $2.160$ & $2.160^{+0.083}_{-0.069}$ & $%
2.03-2.34$ & $1.96-2.41$ \\ \hline
$\delta_{CP}$ & $248.78^\circ$ & $218^{+38^\circ}_{-27^\circ}$ & $%
182^\circ-315^\circ$ & $157^\circ-349^\circ$ \\ \hline
\end{tabular}%
\caption{Model and experimental values for the physical observables of the neutrino sector: neutrino mass squared
splittings, leptonic mixing angles and the leptonic CP phase for the scenario of normal ordering.}
\label{Tab:neutrinofit}
\end{table}
Figure \ref{fig:correlationleptonsector} shows the correlations of the leptonic Dirac CP-violating phase $\delta_{CP}$ with the solar $\sin^2 \theta_{12}$ and with the reactor $\sin^2 \theta_{13}$ mixing parameters as well as the correlations between the leptonic mixing parameters. To obtain these Figures, the lepton sector parameters were randomly generated in a range of values where the neutrino mass squared splittings, leptonic mixing parameters and leptonic Dirac CP violating phase are inside the $3\sigma$ experimentally allowed range. We found a leptonic Dirac CP violating phase in the range $247.5^{\circ}\lesssim\delta_{CP}\lesssim 250.2^{\circ}$, whereas the leptonic mixing parameters are obtained to be in the ranges $0.316\lesssim\sin^2\theta_{12}\lesssim 0.324$, $0.5462\lesssim\sin^2\theta_{23}\lesssim 0.5476$ and $0.0208\lesssim\sin^2\theta_{13}\lesssim 0.0224$.
\newpage

\section{Higgs diphoton decay rate constraints.}

\label{Higgsdiphotonrate}

The decay rate expression for the $h\rightarrow \gamma \gamma$ process is given by \cite{Shifman:1979eb,Gavela:1981ri,Kalyniak:1985ct,Gunion:1989we,Spira:1997dg,Djouadi:2005gj,Marciano:2011gm,Wang:2012gm}:

\begin{align}
\Gamma(h \rightarrow \gamma\gamma) = \dfrac{\alpha_{em}^2 m_h^3}{256 \pi^3 v^2} \biggr| \sum_f a_{hff} N_C Q_f^2 F_{1/2}(\rho_f) + a_{hWW} F_{1}(\rho_W) + a_{hW'W'} F_{1}(\rho_{W'}) + \frac{\lambda_{hH^{\pm}H^{\mp}}v}{2m^{2}_{H^{\pm}}}F_{0}(\rho_{H^{\pm}})\biggr|^2,
\end{align}
where
\begin{align}
    a_{hWW} &= \sin (\beta- \delta ),\\
    a_{hW'W'} &= \cos \delta \sin \gamma, \\
    a_{htt} &= -\frac{\sin \delta}{\sin \beta},\\
    \lambda_{hH^{\pm}H^{\pm}} &= 2 \left(-\lambda _5 \sin (\delta
   ) \sin ^2(\gamma ) \nu
   _{\eta }+\lambda _6
   (-\sin (\delta )) \cos
   ^2(\gamma ) \nu _{\eta
   }+\cos (\delta ) \left(\nu
   _{\rho }\right. \right. \\
   &\quad +\left. \left.
   \left(\left(\lambda _4+\lambda
   _8\right) \sin ^2(\gamma )+2
   \lambda _2 \cos ^2(\gamma
   )\right)+\lambda _8 \sin
   (\gamma ) \cos (\gamma ) \nu
   _{\chi
   }\right)  \right).
   \label{couplingsHdiphoton}
\end{align}

Here $\rho_i$ are the mass ratios $\rho_i= \frac{m_h^2}{4 M_i^2}$ with $M_i=m_f, M_W, M_{W'}$; $\alpha_{em}$ is the fine structure constant; $N_C$ is the color factor ($N_C=1$ for leptons and $N_C=3$ for quarks); and $Q_f$ is the electric charge of the fermion in the loop. From the fermion-loop contributions we only consider the dominant top quark term. 

Furthermore, $F_{1/2}(z)$ and $F_{1}(z)$ are the dimensionless loop factors for spin-$1/2$ and spin-$1$ particles running in the internal lines of the loops. These loop factors take the form:
\begin{align}
    F_{1/2}(z) &= 2(z + (z -1)f(z))z^{-2}, \\
    F_{1}(z) &= -2(2z^2 + 3z + 3(2z-1)f(z))z^{-2}, \\
    F_{0}(z) &= -(z - f(z))z^{-2},
\end{align}
with
\begin{align}
    f(z) = \left\{ \begin{array}{lcc}
             \arcsin^2 \sqrt{2} & \text{for}  & z \leq 1 \\
             \\ -\frac{1}{4}\left(\ln \left(\frac{1+\sqrt{1-z^{-1}}}{1-\sqrt{1-z^{-1}}-i\pi} \right)^2 \right) &  \text{for} & z > 1\\
             \end{array}
   \right.
\end{align}

\begin{figure}[H]
    \begin{subfigure}{.5\linewidth}
        \centering
        \captionsetup{width=0.8\textwidth}
        \includegraphics[scale=.55]{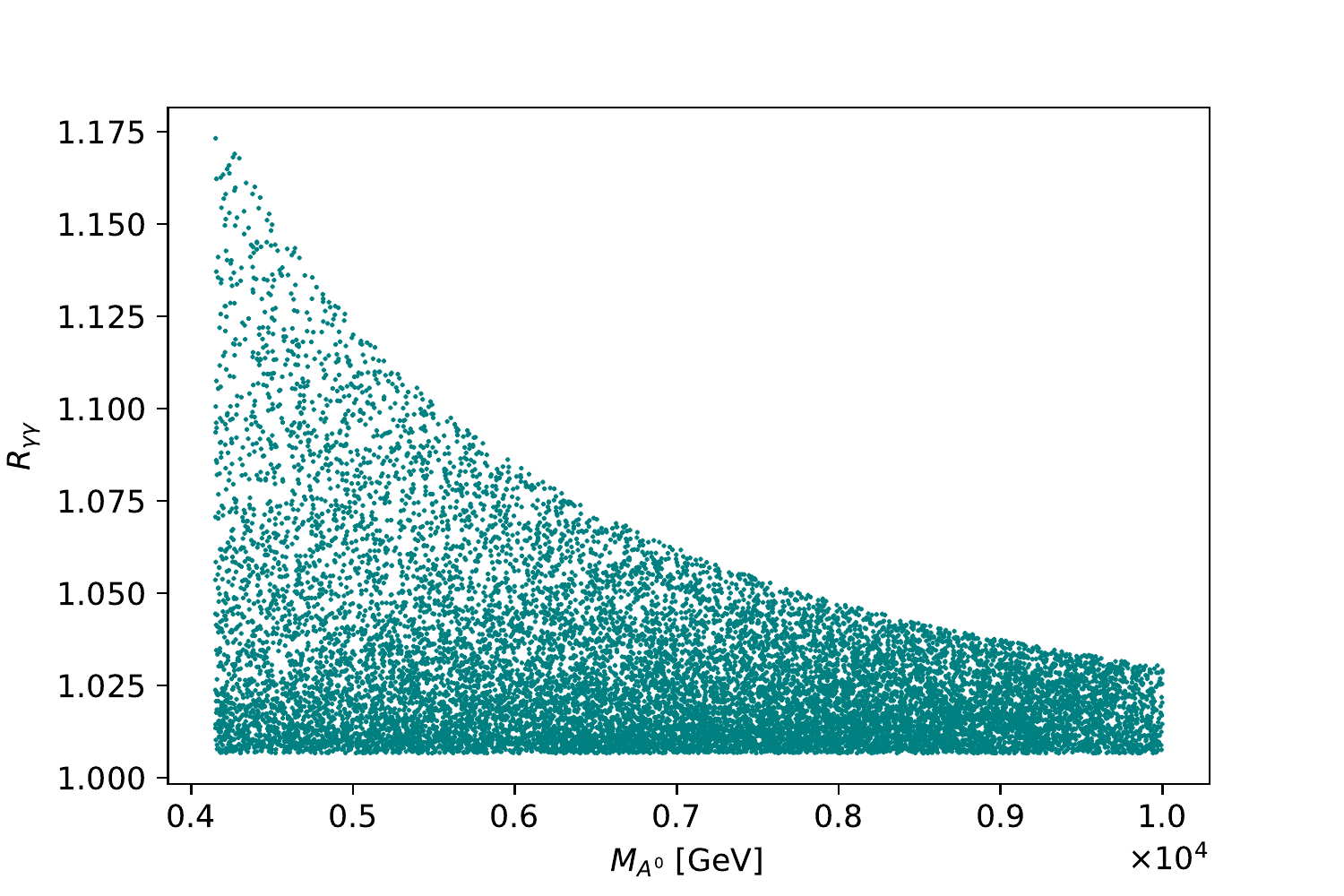}
        \caption{Correlation of the $R_{\gamma\gamma}$ parameter with the CP-odd Higgs mass.}
        \label{fig:sub2}
    \end{subfigure}
    \begin{subfigure}{.5\linewidth}
        \centering
        \captionsetup{width=0.8\textwidth}
        \includegraphics[scale=.55]{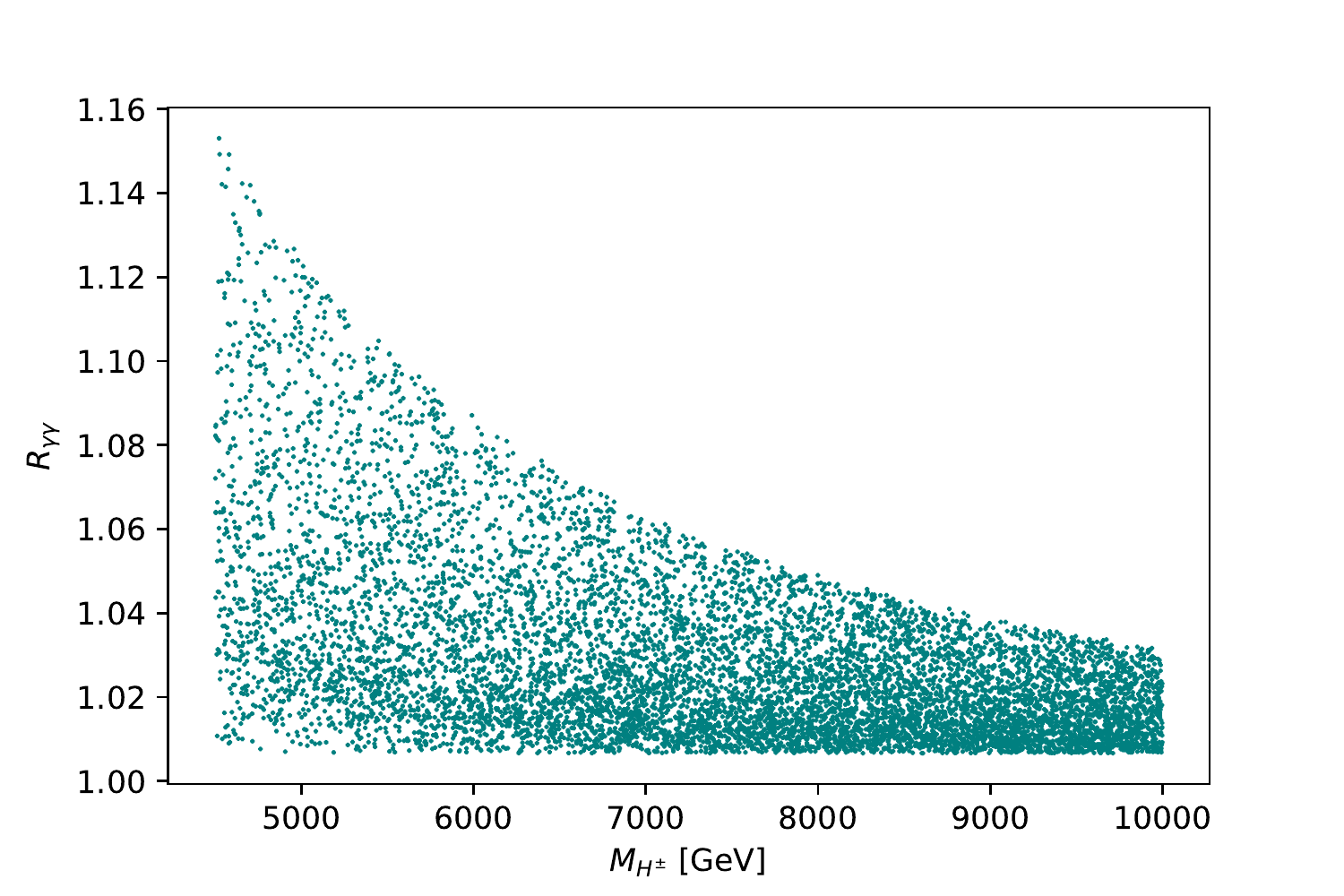}
        \caption{Correlation of the $R_{\gamma\gamma}$ parameter with the charged Higgs boson mass.}
        \label{fig:sub3}
    \end{subfigure}\\[1ex]
    \begin{subfigure}{\linewidth}
        \centering
        \captionsetup{width=0.8\textwidth}
        \includegraphics[scale=.55]{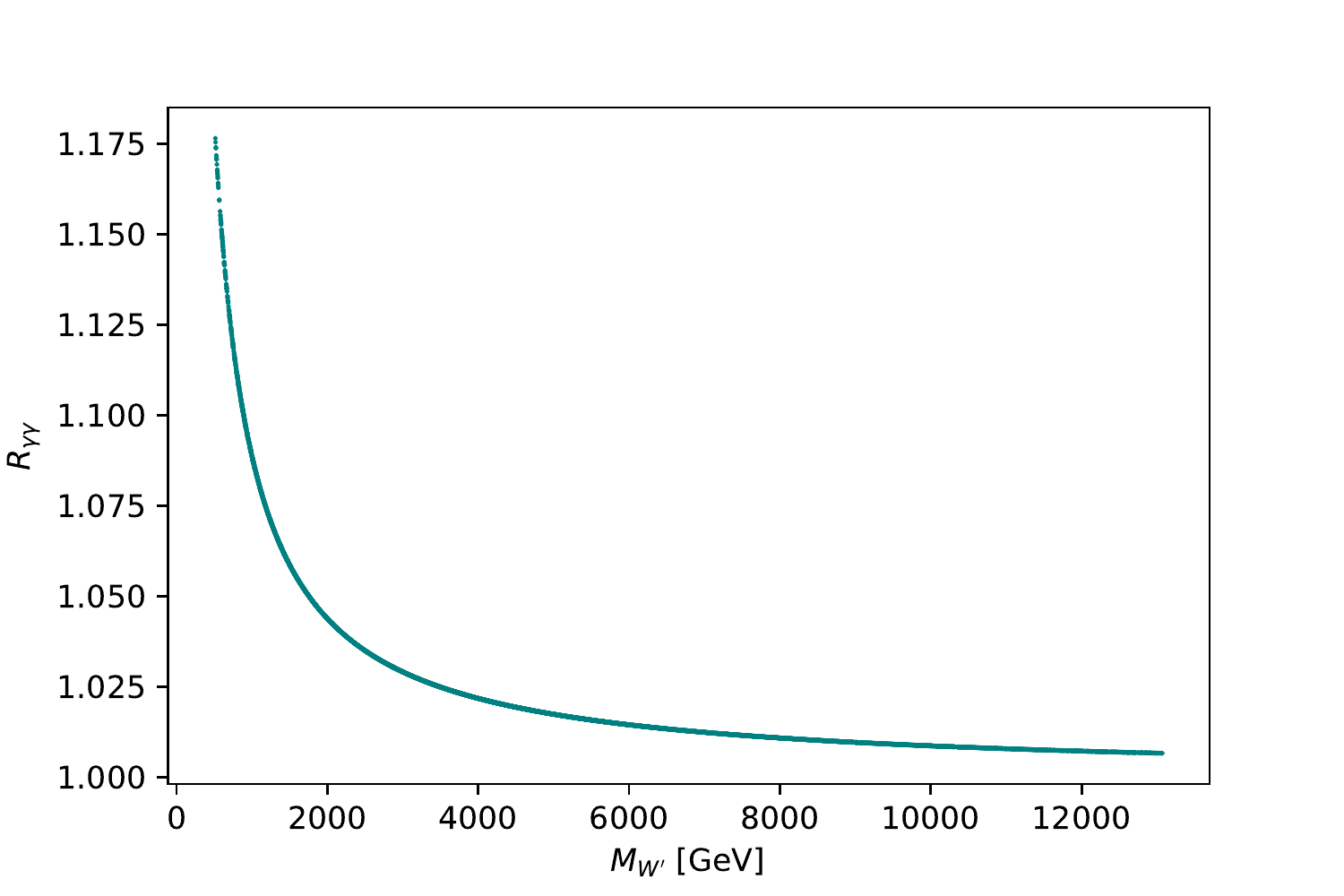}
        \caption{Correlation of the $R_{\gamma\gamma}$ parameter with the $W^\prime$ gauge boson mass.}
        \label{fig:sub4}
    \end{subfigure}
    \caption{Correlations of the $R_{\gamma\gamma}$ parameter with the masses of the CP-odd scalar, charged scalars and $W^\prime$ gauge boson.}
    \label{fig:figRM}
\end{figure}

\begin{figure}[H]
    \centering
    \includegraphics[scale=.55]{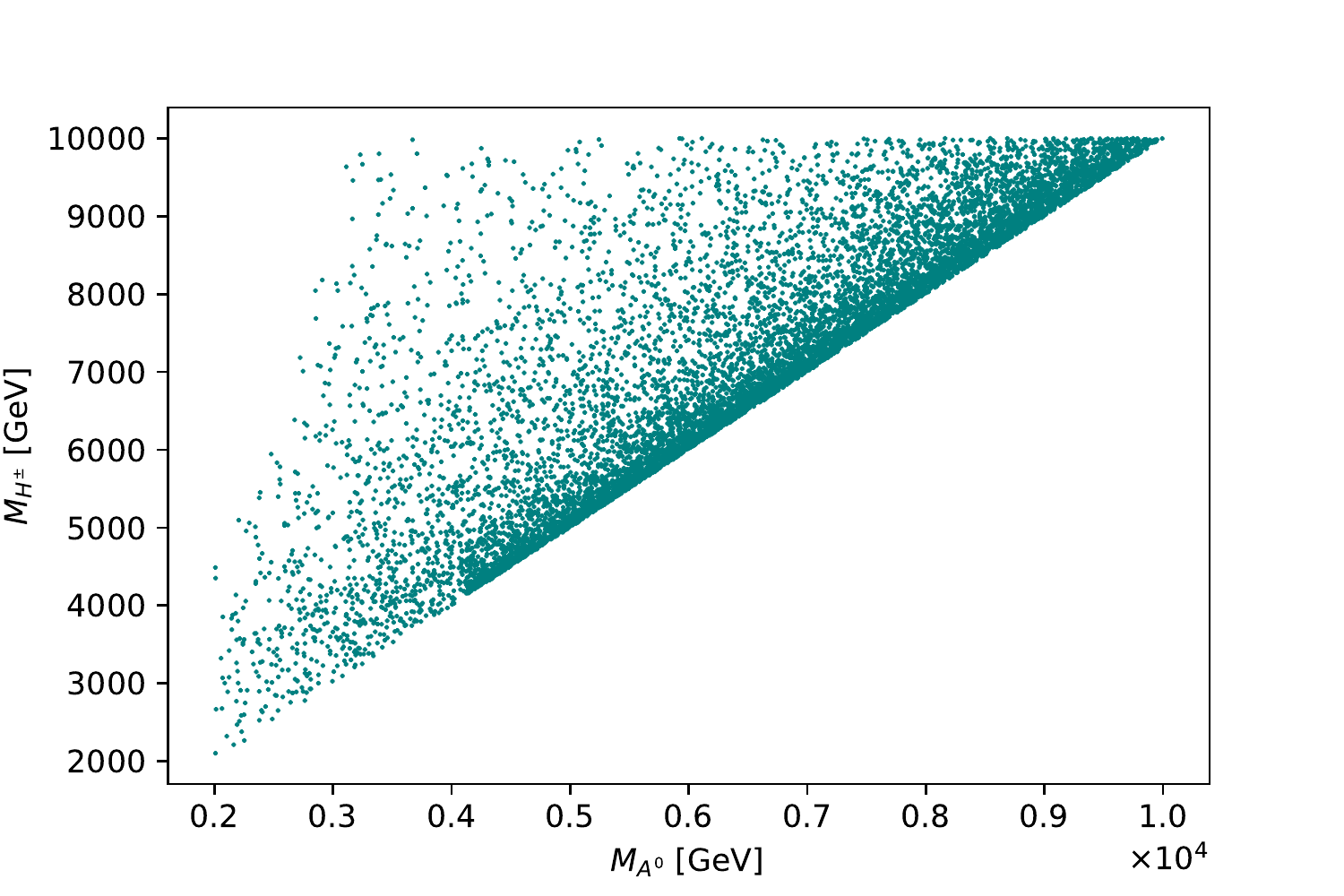}
    \caption{Correlation plot of the CP-odd Higgs mass and the charged Higgs mass.}
    \label{fig:mhca}
\end{figure}

In order to get the constraints on the model parameter space arising from the decay of the $126$ GeV Higgs into a photon pair, the observable $R_{\gamma \gamma}$ is introduced:
\begin{align}
    R_{\gamma \gamma} = \frac{\sigma(pp \to h)\Gamma(h \to \gamma \gamma)}{\sigma(pp \to h)_{SM}\Gamma(h \to \gamma\gamma)_{SM}} \simeq a^{2}_{htt} \frac{\Gamma(h \to \gamma \gamma)}{\Gamma(h \to \gamma \gamma)_{SM}}.
    \label{eqn:hgg}
\end{align}
That observable, which is called the Higgs diphoton signal strength, normalizes the $\gamma \gamma$ signal predicted by our model in relation to the one given by the SM. We have used the same normalization as in Refs. \cite{Campos:2014zaa,Wang:2012gm,Carcamo-Hernandez:2013ypa,Bhattacharyya:2014oka,Fortes:2014dia,Hernandez:2015xka,Hernandez:2015dga,CarcamoHernandez:2017pei}.
%
%
The ratio $R_{\gamma \gamma}$ has been measured by CMS and ATLAS collaborations with the best fit signals \cite{Khachatryan:2014ira,Aad:2014eha}: 
\begin{align}
R^{CMS}_{\gamma \gamma} = 1.14^{+0.26}_{-0.23} \ \text{and} \
R^{ATLAS}_{\gamma \gamma} = 1.17 \pm 0.27.
\label{eqn:rgg}
\end{align}
The best fit result for the ratio $R_{\gamma \gamma}$ is:
\begin{equation}
R_{\gamma \gamma} = 1.0267.
\end{equation}
This value was obtained using the best fit results shown in Table \ref{tab:hgammagamma} and is consistent with the current Higgs diphoton decay rate constraints.
\begin{table}[H]
\centering
\begin{tabular}{c|c}
\textbf{Parameters} & \textbf{Model value} \\ \hline\hline
$M_{h^0}$ & $125.09$ GeV \\ \hline
$M_{H^0}$ & $5319.77$ GeV \\ \hline
$M_{A^0}$ & $5318.3$ GeV \\ \hline
$M_{H^{\pm}}$ & $5503.95$ GeV \\ \hline
$a_{hW^-W^+}$ & $1.0$ \\ \hline
$a_{hW^{\prime }W^{\prime }}$ & $0.0122981$ \\ \hline
$a_{htt}$ & $1.0$ \\ \hline
$\lambda_{hH^{\pm}H^{\pm}}$ & $2525.45$ GeV \\ \hline
\end{tabular}%
\caption{Parameters with $\protect\nu_\protect\eta=173.948$ GeV, $\protect\nu%
_\protect\rho=173.948$ GeV and $\protect\nu_\protect\chi=10$ TeV.}
\label{tab:hgammagamma}
\end{table}
Correlations plots have been obtained to observe the behavior of the $R_{\gamma \gamma}$ parameter as function of the scalar masses and $W^\prime$ gauge boson mass. They are shown in Figure \ref{fig:figRM}. 
These plots were generated using random points in a space in the neighborhood of the best fit values for $f$, $v_{\chi}$ and $\lambda_8$. Figure \ref{fig:sub2} shows that the parameter $R_{\gamma \gamma}$ is strongly restricted by the CP-odd Higgs mass $M_{A^0}$, since the range of allowed values for $R_{\gamma \gamma}$ decreases when the CP odd scalar mass $M_{A^0}$ is increased. The Higgs diphoton signal strength $R_{\gamma \gamma}$ features a similar behavior with the charged scalar mass $M_{H^{\pm}}$, as indicated by Figure \ref{fig:sub3}. Notice that despite the CP odd neutral scalar $A^0$ does not contribute to the Higgs diphoton decay rate, the Higgs diphoton signal strength indirectly depends on $M_{A^0}$ since the parameters $\delta$, $\gamma$ and $\lambda_{hH^{\pm}H^{\pm}}$ (that enter in the Higgs diphoton decay rate) as well as the CP-odd Higgs mass $M_{A^0}$ are functions of $v_{\chi}$. In addition, we have found that the Higgs diphoton signal strength decreases when the $W^\prime$ mass is increased, approaching to 1 when $M_{W{^\prime}}\gtrsim 10$ TeV, as indicated by Figure \ref{fig:sub4}. Furthermore, Figures \ref{fig:sub2}, \ref{fig:sub3} and \ref{fig:sub4} show that our model favors values for the Higgs diphoton decay rate larger than the SM expectation. 
In addition, Figure \ref{fig:mhca} shows that the Higgs diphoton decay rate constraints are fulfilled when $M_{H^{\pm}}\gtrsim M_{A^0}$. Finally, our obtained results for the Higgs diphoton signal strength indicate that the Higgs diphoton decay is a smoking gun signature of our model, whose more precise measurement will be crucial to assess its viability.

\newpage

\section{Heavy scalar production at proton-proton collider}
\label{H1lhc}
In this section we discuss the singly heavy scalar $H_{1}$ production at proton-proton collider. It is worth mentioning that the production mechanism at the LHC of the heavy scalar $H_{1}$ is via gluon fusion, which is a one loop process mediated by the heavy exotic $t^{\prime}$, $j_{1}$ and $j_{2}$ quarks. Consequently, the total $H_{1}$ production cross section in proton proton collisions with center of mass energy $\sqrt{S}$ is given by:
\begin{eqnarray}
\sigma _{pp\rightarrow gg\rightarrow H_{1}}\left( S\right) &=&\frac{\alpha
_{S}^{2}m_{H_{1}}^{2}}{64\pi v_{\chi }^{2}S}\left[ I\left( \frac{m_{H_{1}}^{2}}{m_{t^{\prime}}^{2}}%
\right) +I\left( \frac{m_{H_{1}}^{2}}{m_{j_{1}}^{2}}\right) +I\left( \frac{%
m_{H_{1}}^{2}}{m_{j_{2}}^{2}}\right) \right]  \notag \\
&&\times \int_{\ln \sqrt{\frac{m_{H_{1}}^{2}}{S}}}^{-\ln \sqrt{\frac{%
m_{H_{1}}^{2}}{S}}}f_{p/g}\left( \sqrt{\frac{m_{H_{1}}^{2}}{S}}e^{y},\mu
^{2}\right) f_{p/g}\left( \sqrt{\frac{m_{H_{1}}^{2}}{S}}e^{-y},\mu
^{2}\right) dy,
\end{eqnarray}
where $f_{p/g}\left( x_1,\mu ^2 \right) $ and $f_{p/g}\left(x_2,\mu
^2 \right) $ are the distributions of gluons in the proton which carry
momentum fractions $x_1$ and $x_2$ of the proton, respectively. Furthermore $%
\mu =m_{H_{1}}$ is the factorization scale and $I(z) $ is given
by:
\begin{equation}
I(z)=\int_{0}^{1}dx\int_{0}^{1-x}dy\frac{1-4xy}{1-zxy}.
\label{g1a}
\end{equation}
\begin{figure}[H]
    \centering
    \includegraphics[scale=0.9]{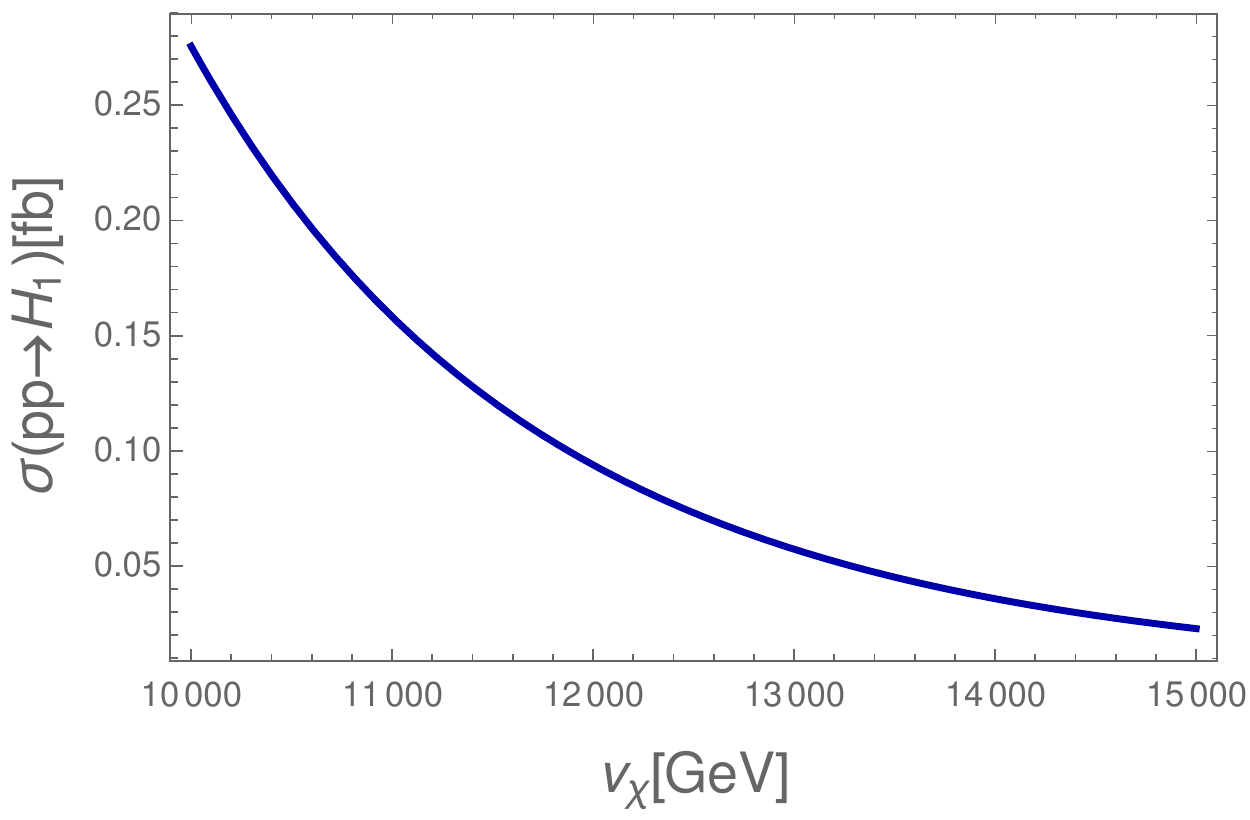}
    \caption{Total cross section for the $H_1$ production via gluon fusion
mechanism at the LHC for $\protect\sqrt{S}=13$ TeV and as a function of the $%
SU(3)_L\times U(1)_X$ symmetry breaking scale $v_{\protect\chi}$ for the scenario described in Eq. (\ref{benchmark}).}
    \label{ggtoH1}
\end{figure}

\begin{figure}[H]
    \centering
    \includegraphics[scale=0.9]{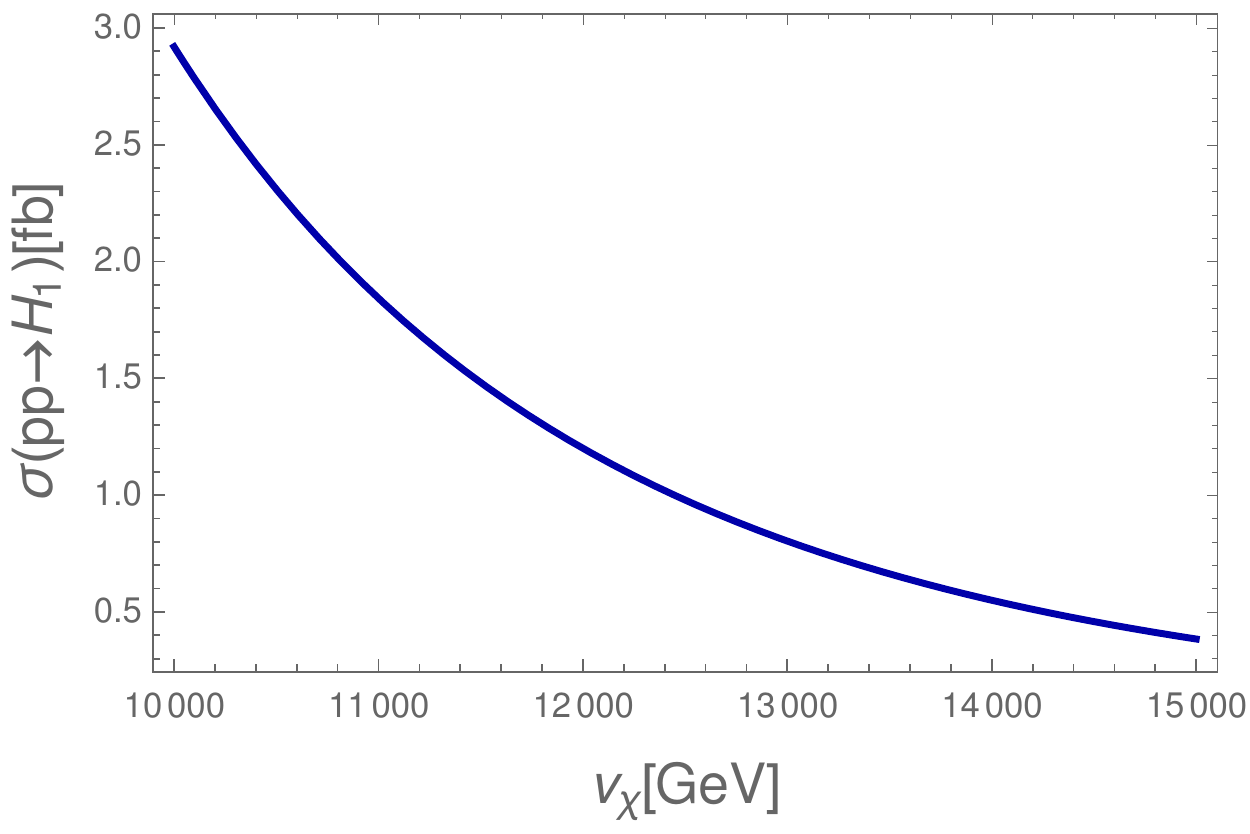}
    \caption{Total cross section for the $H_1$ production via gluon fusion
mechanism at the proposed energy upgrade of the LHC with $\protect\sqrt{S}=28$ TeV as a function of the $SU(3)_L\times U(1)_X$ symmetry breaking scale $v_{\protect\chi}$ for the scenario described in Eq. (\ref{benchmark}).}
    \label{ggtoH1for28TeV}
\end{figure}

\begin{figure}[H]
    \centering
    \includegraphics[scale=0.9]{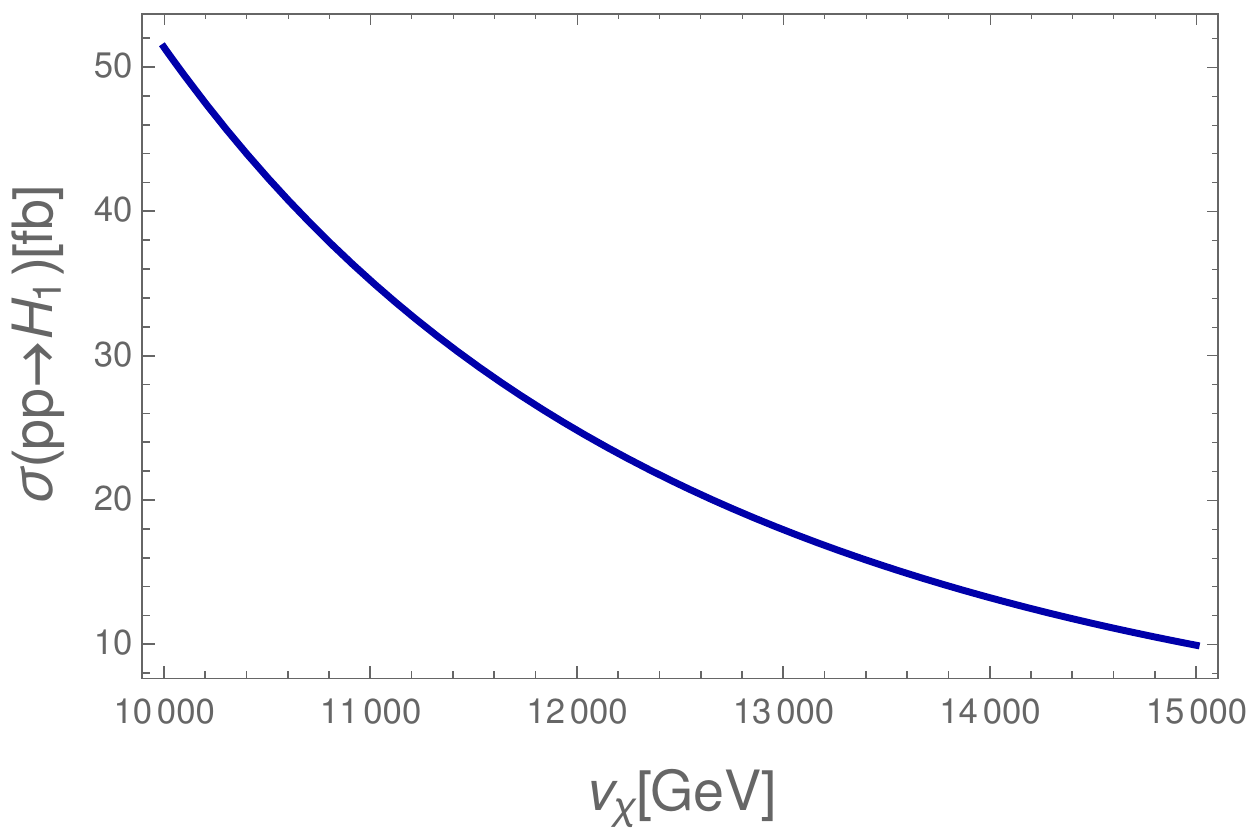}
    \caption{Total cross section for the $H_1$ production via gluon fusion mechanism at a $\protect\sqrt{S}=100$ TeV proton-proton collider as a function of the $SU(3)_L\times U(1)_X$ symmetry breaking scale $v_{\protect\chi}$ for the scenario described in Eq. (\ref{benchmark}).}
    \label{ggtoH1for100TeV}
\end{figure}

Figure \ref{ggtoH1} displays the $H_1$ total production cross section at the
LHC via gluon fusion mechanism for $\sqrt{S}=13$ TeV, as a function of the $%
SU(3)_L\times U(1)_X$ symmetry breaking scale $v_\chi$, which is taken to
range from $10$ TeV up to $15$ TeV, which corresponds to a heavy scalar mass $m_{H_1}$ varying between $1.3$ TeV and $1.9$ TeV. In addition, the exotic quark Yukawa couplings have been taken equal to unity and the scenario described by Eq. (\ref{benchmark}) has been considered in our numerical analysis, Notice that the $SU(3)_L\times U(1)_X$ symmetry breaking scale has been taken larger than $10$ TeV, which corresponds to a $Z^{\prime}$ gauge boson heavier than $4$ TeV, in order to comply with the experimental data on $K$, $D$ and $B$ meson mixings \cite{Huyen:2012uk}. For such region of $H_1$ masses, we find that the total production cross section is found to be $0.28-0.02$ fb. However, at the proposed energy upgrade of the LHC with $\protect\sqrt{S}=28$ TeV, the $H_1$ production cross section is enlarged, reaching values of $2.9-0.4$ fb in the same mass region as indicated by Figure \ref{ggtoH1for28TeV}. Such small values for the $H_1$ production cross section at a proton-proton collider with $\protect\sqrt{S}=13$ TeV and $\protect\sqrt{S}=28$ TeV are small to give rise to a signal for the relevant region of parameter space. However at a $\sqrt{S}=100$ TeV proton-proton collider, there is a significant enhancement of the $H_1$ production cross section, which takes values of $51-10$ fb for $1.3$ TeV $\lesssim$ $m_{H_1}$ $\lesssim$ $1.9$ TeV, as shown in Figure \ref{ggtoH1for100TeV}. Finally, it is worth mentioning that one can safely assume that the heavy $H_1$ scalar after being produced will mainly decay into a pair of SM Higgs bosons, since it is the lightest non SM scalar, as follows from Eq. (\ref{benchmark}) and Table \ref{tab:escalares}. Consequently, an enhancement of the SM Higgs pair production with respect to the SM expectation, will be a smoking gun signature of this model, whose observation will be crucial to assess its viability.

\newpage


\section{Conclusions}
\label{conclusions}
We have constructed a multiscalar singlet extension of the 3-3-1 model with three right handed Majorana neutrinos, consistent with the observed SM fermion mass and mixing pattern. The model incorporates the $S_4$ family symmetry, which is combined with other auxiliary symmetries, thus allowing a viable description of the current SM fermion mass and mixing pattern, which is generated by the spontaneous breaking of the discrete group. The small masses of the light active neutrinos are produced by a linear seesaw mechanism mediated by three Majorana neutrinos. The model provides a successful fit of the physical observables of both quark and lepton sectors. Our model is predictive in the SM quark sector, since it only has 9 effective parameters that allow a successful fit of its 10 observables, i.e., the 6 SM quark masses, the 3 quark mixing parameters and the CP violating phase. In addition, we have found that the SM quark sector of our model has a particular scenario, which is inspired by naturalness arguments and has only 6 effective parameters that allows to successfully reproduce the experimental values of the ten SM quark sector observables. Furthermore, we have also shown that the proposed model successfully accommodates the current Higgs diphoton decay rate constraints provided that the charged Higgs bosons are a bit heavier than the CP odd neutral Higgs boson $A^0$. In addition, we have found that it favors a Higgs diphoton decay rate larger than the SM expectation. Finally, we have also discussed the single production of the heavy scalar $H_1$ associated with the spontaneous breaking of the $SU(3)_C\times U(1)_X$ symmetry, at a proton-proton collider, via gluon fusion mechanism. We have considered the cases where the center of mass energy takes the values of $\protect\sqrt{S}=13$ TeV, $\protect\sqrt{S}=28$ TeV and $\protect\sqrt{S}=100$ TeV. For the first two cases corresponding to the current LHC center of mass energy and the proposed energy upgrade of the LHC, respectively, we have found that the $H_1$ production cross sections are small to give rise to a signal for the relevant region of parameter space. However, in a future $\sqrt{S}=100$ TeV proton-proton collider, the $H_1$ production cross section is significantly enhanced reaching values between $51$ fb and $10$ fb, for the mass range $1.3$ TeV $\lesssim$ $m_{H_1}$$\lesssim$ $1.9$ TeV.

\section*{Acknowledgments}
This research has received funding from Fondecyt (Chile), Grants No.~1170803, CONICYT PIA/Basal FB0821 and the Programa de Incentivos a la Iniciación Científica (PIIC) from USM. A.E.C.H is very grateful to Professor Hoang Ngoc Long for hospitality at the Institute of Physics, Vietnam Academy of Science and Technology, where this work was finished.

\appendix

\section{The $S_{4}$ discrete group} \label{appen:s4}

\label{S4}The $S_{4}$ is the smallest non abelian group having doublet,
triplet and singlet irreducible representations. $S_{4}$ is the group of
permutations of four objects, which includes five irreducible
representations, i.e., $\mathbf{1,1^{\prime },2,3,3^{\prime }}$ fulfilling
the following tensor product rules \cite{Ishimori:2010au}:
\begin{align}
& \mathbf{3}\otimes \mathbf{3}=\mathbf{1}\oplus \mathbf{2}\oplus \mathbf{3}%
\oplus \mathbf{3^{\prime }},\qquad \mathbf{3^{\prime }}\otimes \mathbf{%
3^{\prime }}=\mathbf{1}\oplus \mathbf{2}\oplus \mathbf{3}\oplus \mathbf{%
3^{\prime }},\qquad \mathbf{3}\otimes \mathbf{3^{\prime }}=\mathbf{1^{\prime
}}\oplus \mathbf{2}\oplus \mathbf{3}\oplus \mathbf{3^{\prime }}, \\
& \mathbf{2}\otimes \mathbf{2}=\mathbf{1}\oplus \mathbf{1^{\prime }}\oplus 
\mathbf{2},\qquad \mathbf{2}\otimes \mathbf{3}=\mathbf{3}\oplus \mathbf{%
3^{\prime }},\qquad \mathbf{2}\otimes \mathbf{3^{\prime }}=\mathbf{3^{\prime
}}\oplus \mathbf{3}, \\
& \mathbf{3}\otimes \mathbf{1^{\prime }}=\mathbf{3^{\prime }},\qquad \mathbf{%
3^{\prime }}\otimes \mathbf{1^{\prime }}=\mathbf{3},\qquad \mathbf{2}\otimes 
\mathbf{1^{\prime }}=\mathbf{2}.
\end{align}%
Explicitly, the basis used in this paper corresponds to Ref. \cite%
{Ishimori:2010au} and results in 
\begin{equation}
(\mathbf{A})_{\mathbf{3}}\times (\mathbf{B})_{\mathbf{3}}=(\mathbf{A}\cdot 
\mathbf{B})_{\mathbf{1}}+\left( 
\begin{array}{c}
\mathbf{A}\cdot \Sigma \cdot \mathbf{B} \\ 
\mathbf{A}\cdot \Sigma ^{\ast }\cdot \mathbf{B}%
\end{array}%
\right) _{\mathbf{2}}+\left( 
\begin{array}{c}
\{A_{y}B_{z}\} \\ 
\{A_{z}B_{x}\} \\ 
\{A_{x}B_{y}\}%
\end{array}%
\right) _{\mathbf{3}}+\left( 
\begin{array}{c}
\left[ A_{y}B_{z}\right] \\ 
\left[ A_{z}B_{x}\right] \\ 
\left[ A_{x}B_{y}\right]%
\end{array}%
\right) _{\mathbf{3^{\prime }}},
\end{equation}%
\begin{equation}
(\mathbf{A})_{\mathbf{3^{\prime }}}\times (\mathbf{B})_{\mathbf{3^{\prime }}%
}=(\mathbf{A}\cdot \mathbf{B})_{\mathbf{1}}+\left( 
\begin{array}{c}
\mathbf{A}\cdot \Sigma \cdot \mathbf{B} \\ 
\mathbf{A}\cdot \Sigma ^{\ast }\cdot \mathbf{B}%
\end{array}%
\right) _{\mathbf{2}}+\left( 
\begin{array}{c}
\{A_{y}B_{z}\} \\ 
\{A_{z}B_{x}\} \\ 
\{A_{x}B_{y}\}%
\end{array}%
\right) _{\mathbf{3}}+\left( 
\begin{array}{c}
\left[ A_{y}B_{z}\right] \\ 
\left[ A_{z}B_{x}\right] \\ 
\left[ A_{x}B_{y}\right]%
\end{array}%
\right) _{\mathbf{3^{\prime }}},
\end{equation}%
\begin{equation}
(\mathbf{A})_{\mathbf{3}}\times (\mathbf{B})_{\mathbf{3^{\prime }}}=(\mathbf{%
A}\cdot \mathbf{B})_{\mathbf{1^{\prime }}}+\left( 
\begin{array}{c}
\mathbf{A}\cdot \Sigma \cdot \mathbf{B} \\ 
-\mathbf{A}\cdot \Sigma ^{\ast }\cdot \mathbf{B}%
\end{array}%
\right) _{\mathbf{2}}+\left( 
\begin{array}{c}
\{A_{y}B_{z}\} \\ 
\{A_{z}B_{x}\} \\ 
\{A_{x}B_{y}\}%
\end{array}%
\right) _{\mathbf{3^{\prime }}}+\left( 
\begin{array}{c}
\left[ A_{y}B_{z}\right] \\ 
\left[ A_{z}B_{x}\right] \\ 
\left[ A_{x}B_{y}\right]%
\end{array}%
\right) _{\mathbf{3}},
\end{equation}%
\begin{equation}
(\mathbf{A})_{\mathbf{2}}\times (\mathbf{B})_{\mathbf{2}}=\{A_{x}B_{y}\}_{%
\mathbf{1}}+\left[ A_{x}B_{y}\right] _{\mathbf{1^{\prime }}}+\left( 
\begin{array}{c}
A_{y}B_{y} \\ 
A_{x}B_{x}%
\end{array}%
\right) _{\mathbf{2}},
\end{equation}%
\begin{equation}
\left( 
\begin{array}{c}
A_{x} \\ 
A_{y}%
\end{array}%
\right) _{\mathbf{2}}\times \left( 
\begin{array}{c}
B_{x} \\ 
B_{y} \\ 
B_{z}%
\end{array}%
\right) _{\mathbf{3}}=\left( 
\begin{array}{c}
(A_{x}+A_{y})B_{x} \\ 
(\omega ^{2}A_{x}+\omega A_{y})B_{y} \\ 
(\omega A_{x}+\omega ^{2}A_{y})B_{z}%
\end{array}%
\right) _{\mathbf{3}}+\left( 
\begin{array}{c}
(A_{x}-A_{y})B_{x} \\ 
(\omega ^{2}A_{x}-\omega A_{y})B_{y} \\ 
(\omega A_{x}-\omega ^{2}A_{y})B_{z}%
\end{array}%
\right) _{\mathbf{3}^{\prime }},
\end{equation}%
\begin{equation}
\left( 
\begin{array}{c}
A_{x} \\ 
A_{y}%
\end{array}%
\right) _{\mathbf{2}}\times \left( 
\begin{array}{c}
B_{x} \\ 
B_{y} \\ 
B_{z}%
\end{array}%
\right) _{\mathbf{3}^{\prime }}=\left( 
\begin{array}{c}
(A_{x}+A_{y})B_{x} \\ 
(\omega ^{2}A_{x}+\omega A_{y})B_{y} \\ 
(\omega A_{x}+\omega ^{2}A_{y})B_{z}%
\end{array}%
\right) _{\mathbf{3}^{\prime }}+\left( 
\begin{array}{c}
(A_{x}-A_{y})B_{x} \\ 
(\omega ^{2}A_{x}-\omega A_{y})B_{y} \\ 
(\omega A_{x}-\omega ^{2}A_{y})B_{z}%
\end{array}%
\right) _{\mathbf{3}},
\end{equation}%
with 
\begin{align}
\mathbf{A}\cdot \mathbf{B}& =A_{x}B_{x}+A_{y}B_{y}+A_{z}B_{z},  \notag \\
\{A_{x}B_{y}\}& =A_{x}B_{y}+A_{y}B_{x},  \notag \\
\left[ A_{x}B_{y}\right] & =A_{x}B_{y}-A_{y}B_{x},  \notag \\
\mathbf{A}\cdot \Sigma \cdot \mathbf{B}& =A_{x}B_{x}+\omega
A_{y}B_{y}+\omega ^{2}A_{z}B_{z},  \notag \\
\mathbf{A}\cdot \Sigma ^{\ast }\cdot \mathbf{B}& =A_{x}B_{x}+\omega
^{2}A_{y}B_{y}+\omega A_{z}B_{z},
\end{align}%
where $\omega =e^{2\pi i/3}$ is a complex square root of unity.

\section{The scalar potential for a $S_{4}$ doublet} \label{scalarpotentialS4doublet}

The scalar potential for a $S_{4}$ doublet $\Delta$ is given by:
\begin{align}\label{eqn:scalpotdelta}
    V =& -\mu^2_{\Delta} (\Delta \Delta^{*})_{\textbf{1}} + \kappa_1 (\Delta \Delta^{*})_{\textbf{1}} (\Delta \Delta^{*})_{\textbf{1}} + \kappa_2 (\Delta \Delta^{*})_{\textbf{1'}} (\Delta \Delta^{*})_{\textbf{1'}} + \kappa_3 (\Delta \Delta^{*})_{\textbf{2}} (\Delta \Delta^{*})_{\textbf{2}} + h.c.
\end{align}
This expression has four free parameters: one bilinear and three quartic couplings. The $\mu_{\Delta}$ parameter can be written as a function of the other three parameters by using the scalar potential minimization condition:
\begin{align}
    \frac{\partial \langle V(\Delta) \rangle}{\partial v_{\Delta}} =& 16 \kappa_1 v_{\Delta}^3 + 8 \kappa_3 v_{\Delta}^3- v_{\Delta}\mu_{\Delta}^2\notag\\
    =&0.
\end{align}
Solving the leading equation for $\mu^{2}_{\Delta}$ yields the following relation:
\begin{equation}
    \mu^{2}_{\Delta} = 8(2\kappa_{1} + \kappa_{3})v^{2}_{\Delta}.
\end{equation}

This result indicates that the VEV pattern of the $S_4$ doublet $\Delta$ given in Eq. \eqref{eqn:VEV}, is consistent with a global minimum of the scalar potential of Eq. \eqref{eqn:scalpotdelta} for a large region of parameter space. The previously described procedure can be used to show that the VEV patterns of the remaining $S_4$ doublets of the model are also consistent with the minimization conditions of the scalar potential.

\section{The scalar potential for a $S_{4}$ triplet} \label{scalarpotentialS4triplet}

The scalar potential for a $S_4$ triplet $S$ has six free parameters: one bilinear and four quartic couplings, as indicated by the relation:
\begin{align}\label{eqn:scalpotphi}
   V =& -\mu^{2}_{S} (SS^{*})_{\textbf{1}} + \kappa_1 (SS^{*})_{\textbf{1}}(SS^{*})_{\textbf{1}} + \kappa_2(SS^{*})_{\textbf{3}}(SS^{*})_{\textbf{3}} + \kappa_3 (SS^{*})_{\textbf{3`}}(SS^{*})_{\textbf{3`}} + \kappa_4 (SS^{*})_{\textbf{2}} (SS^{*})_{\textbf{2}} + h.c.
\end{align}

Its minimization equation allows us to express the $\mu_{S}$ parameter as follows:
\begin{align}
    \frac{\partial \langle V(S) \rangle}{\partial v_{S}} = & 36 \kappa _1 v _{S}^3+48 \kappa _2 v _{S}^3+4 \kappa _4 \left(2 e^{\frac{2 i \pi
   }{3}} v _{S}+2 e^{-\frac{2 i \pi }{3}} v _{S}+2 v _{S}\right)
   \left(e^{\frac{2 i \pi }{3}} v _{S}^2+e^{-\frac{2 i \pi }{3}} v _{S}^2+v_{S}^2\right)-6 \mu^{2}_{S} v _{S}\notag\\ &=0.
   \label{minimizationconditionS4triplet}
\end{align}
%
%
Here we consider the phase $\omega = e^{2\pi i/3}$ arising in the tensor product of $S_4$ scalar triplets. Then, we find the following relation for the $\mu_{S}^{2}$ parameter:
\begin{equation}
\mu_{S}^{2} = 2\left(3 \kappa _1+4 \kappa _2\right)v_{S}^2.
\end{equation}

This shows that the VEV configuration of the $S_4$ triplet $S$ given in Eq. \eqref{eqn:VEV}, is in accordance with the scalar potential minimization condition of Eq. \eqref{minimizationconditionS4triplet}. The remaining $S_4$ triplets of the model are also consistent with the minimization conditions of the scalar potential and this can be proved by using the same procedure described before.


\begin{thebibliography}{9}
\bibitem{Georgi:1978bv} 
  H.~Georgi and A.~Pais,
  Phys.\ Rev.\ D {\bf 19}, 2746 (1979).
  doi:10.1103/PhysRevD.19.2746



\bibitem{Valle:1983dk} 
  J.~W.~F.~Valle and M.~Singer,
  Phys.\ Rev.\ D {\bf 28}, 540 (1983).
  doi:10.1103/PhysRevD.28.540



\bibitem{Pisano:1991ee} 
  F.~Pisano and V.~Pleitez,
  Phys.\ Rev.\ D {\bf 46}, 410 (1992)
  doi:10.1103/PhysRevD.46.410
  [hep-ph/9206242].



\bibitem{Foot:1992rh} 
  R.~Foot, O.~F.~Hernandez, F.~Pisano and V.~Pleitez,
  Phys.\ Rev.\ D {\bf 47}, 4158 (1993)
  doi:10.1103/PhysRevD.47.4158
  [hep-ph/9207264].



\bibitem{Frampton:1992wt} 
  P.~H.~Frampton,
  Phys.\ Rev.\ Lett.\  {\bf 69}, 2889 (1992).
  doi:10.1103/PhysRevLett.69.2889



\bibitem{Hoang:1996gi} 
  H.~N.~Long,
  Phys.\ Rev.\ D {\bf 54}, 4691 (1996)
  doi:10.1103/PhysRevD.54.4691
  [hep-ph/9607439].



\bibitem{Hoang:1995vq} 
  H.~N.~Long,
  Phys.\ Rev.\ D {\bf 53}, 437 (1996)
  doi:10.1103/PhysRevD.53.437
  [hep-ph/9504274].



\bibitem{Foot:1994ym} 
  R.~Foot, H.~N.~Long and T.~A.~Tran,
  Phys.\ Rev.\ D {\bf 50}, no. 1, R34 (1994)
  doi:10.1103/PhysRevD.50.R34
  [hep-ph/9402243].



\bibitem{CarcamoHernandez:2005ka} 
  A.~E.~Carcamo Hernandez, R.~Martinez and F.~Ochoa,
  Phys.\ Rev.\ D {\bf 73}, 035007 (2006)
  doi:10.1103/PhysRevD.73.035007
  [hep-ph/0510421].



\bibitem{Dong:2010zu} 
  P.~V.~Dong, H.~N.~Long, D.~V.~Soa and V.~V.~Vien,
  Eur.\ Phys.\ J.\ C {\bf 71}, 1544 (2011)
  doi:10.1140/epjc/s10052-011-1544-2
  [arXiv:1009.2328 [hep-ph]].



\bibitem{Dong:2010gk} 
  P.~V.~Dong, L.~T.~Hue, H.~N.~Long and D.~V.~Soa,
  Phys.\ Rev.\ D {\bf 81}, 053004 (2010)
  doi:10.1103/PhysRevD.81.053004
  [arXiv:1001.4625 [hep-ph]].



\bibitem{Dong:2011vb} 
  P.~V.~Dong, H.~N.~Long, C.~H.~Nam and V.~V.~Vien,
  Phys.\ Rev.\ D {\bf 85}, 053001 (2012)
  doi:10.1103/PhysRevD.85.053001
  [arXiv:1111.6360 [hep-ph]].



\bibitem{Benavides:2010zw} 
  R.~H.~Benavides, W.~A.~Ponce and Y.~Giraldo,
  Phys.\ Rev.\ D {\bf 82}, 013004 (2010)
  doi:10.1103/PhysRevD.82.013004
  [arXiv:1006.3248 [hep-ph]].



\bibitem{Dong:2012bf} 
  P.~V.~Dong, H.~N.~Long and H.~T.~Hung,
  Phys.\ Rev.\ D {\bf 86}, 033002 (2012)
  doi:10.1103/PhysRevD.86.033002
  [arXiv:1205.5648 [hep-ph]].



\bibitem{Huong:2012pg} 
  D.~T.~Huong, L.~T.~Hue, M.~C.~Rodriguez and H.~N.~Long,
  Nucl.\ Phys.\ B {\bf 870}, 293 (2013)
  doi:10.1016/j.nuclphysb.2013.01.016
  [arXiv:1210.6776 [hep-ph]].



\bibitem{Giang:2012vs} 
  P.~T.~Giang, L.~T.~Hue, D.~T.~Huong and H.~N.~Long,
  Nucl.\ Phys.\ B {\bf 864}, 85 (2012)
  doi:10.1016/j.nuclphysb.2012.06.008
  [arXiv:1204.2902 [hep-ph]].



\bibitem{Binh:2013axa} 
  D.~T.~Binh, L.~T.~Hue, D.~T.~Huong and H.~N.~Long,
  Eur.\ Phys.\ J.\ C {\bf 74}, no. 5, 2851 (2014)
  doi:10.1140/epjc/s10052-014-2851-1
  [arXiv:1308.3085 [hep-ph]].



\bibitem{Hernandez:2013mcf} 
  A.~E.~Carcamo Hernandez, R.~Martinez and F.~Ochoa,
  Phys.\ Rev.\ D {\bf 87}, no. 7, 075009 (2013)
  doi:10.1103/PhysRevD.87.075009
  [arXiv:1302.1757 [hep-ph]].



\bibitem{Hernandez:2013hea} 
  A.~E.~Cárcamo Hernández, R.~Martinez and F.~Ochoa,
  Eur.\ Phys.\ J.\ C {\bf 76}, no. 11, 634 (2016)
  doi:10.1140/epjc/s10052-016-4480-3
  [arXiv:1309.6567 [hep-ph]].



\bibitem{Hernandez:2014vta} 
  A.~E.~Cárcamo Hernández, R.~Martinez and J.~Nisperuza,
  Eur.\ Phys.\ J.\ C {\bf 75}, no. 2, 72 (2015)
  doi:10.1140/epjc/s10052-015-3278-z
  [arXiv:1401.0937 [hep-ph]].



\bibitem{Hernandez:2014lpa} 
  A.~E.~Cárcamo Hernández, E.~Cataño Mur and R.~Martinez,
  Phys.\ Rev.\ D {\bf 90}, no. 7, 073001 (2014)
  doi:10.1103/PhysRevD.90.073001
  [arXiv:1407.5217 [hep-ph]].



\bibitem{Kelso:2014qka} 
  C.~Kelso, H.~N.~Long, R.~Martinez and F.~S.~Queiroz,
  Phys.\ Rev.\ D {\bf 90}, no. 11, 113011 (2014)
  doi:10.1103/PhysRevD.90.113011
  [arXiv:1408.6203 [hep-ph]].



\bibitem{Vien:2014gza} 
  V.~V.~Vien and H.~N.~Long,
  JHEP {\bf 1404}, 133 (2014)
  doi:10.1007/JHEP04(2014)133
  [arXiv:1402.1256 [hep-ph]].



\bibitem{Phong:2014ofa} 
  V.~Q.~Phong, H.~N.~Long, V.~T.~Van and L.~H.~Minh,
  Eur.\ Phys.\ J.\ C {\bf 75}, no. 7, 342 (2015)
  doi:10.1140/epjc/s10052-015-3550-2
  [arXiv:1409.0750 [hep-ph]].



\bibitem{Phong:2014yca} 
  V.~Q.~Phong, H.~N.~Long, V.~T.~Van and N.~C.~Thanh,
  Phys.\ Rev.\ D {\bf 90}, no. 8, 085019 (2014)
  doi:10.1103/PhysRevD.90.085019
  [arXiv:1408.5657 [hep-ph]].



\bibitem{Boucenna:2014ela} 
  S.~M.~Boucenna, S.~Morisi and J.~W.~F.~Valle,
  Phys.\ Rev.\ D {\bf 90}, no. 1, 013005 (2014)
  doi:10.1103/PhysRevD.90.013005
  [arXiv:1405.2332 [hep-ph]].



\bibitem{DeConto:2015eia} 
  G.~De Conto, A.~C.~B.~Machado and V.~Pleitez,
  Phys.\ Rev.\ D {\bf 92}, no. 7, 075031 (2015)
  doi:10.1103/PhysRevD.92.075031
  [arXiv:1505.01343 [hep-ph]].



\bibitem{Boucenna:2015zwa} 
  S.~M.~Boucenna, J.~W.~F.~Valle and A.~Vicente,
  Phys.\ Rev.\ D {\bf 92}, no. 5, 053001 (2015)
  doi:10.1103/PhysRevD.92.053001
  [arXiv:1502.07546 [hep-ph]].



\bibitem{Boucenna:2015pav} 
  S.~M.~Boucenna, S.~Morisi and A.~Vicente,
  Phys.\ Rev.\ D {\bf 93}, no. 11, 115008 (2016)
  doi:10.1103/PhysRevD.93.115008
  [arXiv:1512.06878 [hep-ph]].



\bibitem{Benavides:2015afa} 
  R.~H.~Benavides, L.~N.~Epele, H.~Fanchiotti, C.~G.~Canal and W.~A.~Ponce,
  Adv.\ High Energy Phys.\  {\bf 2015}, 813129 (2015)
  doi:10.1155/2015/813129
  [arXiv:1503.01686 [hep-ph]].



\bibitem{Hernandez:2015tna} 
  A.~E.~Cárcamo Hernández and R.~Martinez,
  Nucl.\ Phys.\ B {\bf 905}, 337 (2016)
  doi:10.1016/j.nuclphysb.2016.02.025
  [arXiv:1501.05937 [hep-ph]].



\bibitem{Hue:2015fbb} 
  L.~T.~Hue, H.~N.~Long, T.~T.~Thuc and T.~Phong Nguyen,
  Nucl.\ Phys.\ B {\bf 907}, 37 (2016)
  doi:10.1016/j.nuclphysb.2016.03.034
  [arXiv:1512.03266 [hep-ph]].



\bibitem{Hernandez:2015ywg} 
  A.~E.~C.~Hernández and I.~Nišandžić,
  Eur.\ Phys.\ J.\ C {\bf 76}, no. 7, 380 (2016)
  doi:10.1140/epjc/s10052-016-4230-6
  [arXiv:1512.07165 [hep-ph]].



\bibitem{Fonseca:2016tbn} 
  R.~M.~Fonseca and M.~Hirsch,
  JHEP {\bf 1608}, 003 (2016)
  doi:10.1007/JHEP08(2016)003
  [arXiv:1606.01109 [hep-ph]].



\bibitem{Vien:2016tmh} 
  V.~V.~Vien, A.~E.~Cárcamo Hernández and H.~N.~Long,
  Nucl.\ Phys.\ B {\bf 913}, 792 (2016)
  doi:10.1016/j.nuclphysb.2016.10.010
  [arXiv:1601.03300 [hep-ph]].



\bibitem{Hernandez:2016eod} 
  A.~E.~Cárcamo Hernández, H.~N.~Long and V.~V.~Vien,
  Eur.\ Phys.\ J.\ C {\bf 76}, no. 5, 242 (2016)
  doi:10.1140/epjc/s10052-016-4074-0
  [arXiv:1601.05062 [hep-ph]].



\bibitem{Fonseca:2016xsy} 
  R.~M.~Fonseca and M.~Hirsch,
  Phys.\ Rev.\ D {\bf 94}, no. 11, 115003 (2016)
  doi:10.1103/PhysRevD.94.115003
  [arXiv:1607.06328 [hep-ph]].



\bibitem{Deppisch:2016jzl} 
  F.~F.~Deppisch, C.~Hati, S.~Patra, U.~Sarkar and J.~W.~F.~Valle,
  Phys.\ Lett.\ B {\bf 762}, 432 (2016)
  doi:10.1016/j.physletb.2016.10.002
  [arXiv:1608.05334 [hep-ph]].



\bibitem{Reig:2016ewy} 
  M.~Reig, J.~W.~F.~Valle and C.~A.~Vaquera-Araujo,
  Phys.\ Rev.\ D {\bf 94}, no. 3, 033012 (2016)
  doi:10.1103/PhysRevD.94.033012
  [arXiv:1606.08499 [hep-ph]].



\bibitem{CarcamoHernandez:2017cwi} 
  A.~E.~Cárcamo Hernández, S.~Kovalenko, H.~N.~Long and I.~Schmidt,
  JHEP {\bf 1807}, 144 (2018)
  doi:10.1007/JHEP07(2018)144
  [arXiv:1705.09169 [hep-ph]].



\bibitem{CarcamoHernandez:2017kra} 
  A.~E.~Cárcamo Hernández and H.~N.~Long,
  J.\ Phys.\ G {\bf 45}, no. 4, 045001 (2018)
  doi:10.1088/1361-6471/aaace7
  [arXiv:1705.05246 [hep-ph]].



\bibitem{Hati:2017aez} 
  C.~Hati, S.~Patra, M.~Reig, J.~W.~F.~Valle and C.~A.~Vaquera-Araujo,
  Phys.\ Rev.\ D {\bf 96}, no. 1, 015004 (2017)
  doi:10.1103/PhysRevD.96.015004
  [arXiv:1703.09647 [hep-ph]].



\bibitem{Barreto:2017xix} 
  E.~R.~Barreto, A.~G.~Dias, J.~Leite, C.~C.~Nishi, R.~L.~N.~Oliveira and W.~C.~Vieira,
  Phys.\ Rev.\ D {\bf 97}, no. 5, 055047 (2018)
  doi:10.1103/PhysRevD.97.055047
  [arXiv:1709.09946 [hep-ph]].



\bibitem{CarcamoHernandez:2018iel} 
  A.~E.~Cárcamo Hernández, H.~N.~Long and V.~V.~Vien,
  Eur.\ Phys.\ J.\ C {\bf 78}, no. 10, 804 (2018)
  doi:10.1140/epjc/s10052-018-6284-0
  [arXiv:1803.01636 [hep-ph]].



\bibitem{Vien:2018otl} 
  V.~V.~Vien, H.~N.~Long and A.~E.~Cárcamo Hernández,
  Mod.\ Phys.\ Lett.\ A {\bf 34}, no. 01, 1950005 (2019)
  doi:10.1142/S0217732319500056
  [arXiv:1812.07263 [hep-ph]].



\bibitem{Dias:2018ddy} 
  A.~G.~Dias, J.~Leite, D.~D.~Lopes and C.~C.~Nishi,
  Phys.\ Rev.\ D {\bf 98}, no. 11, 115017 (2018)
  doi:10.1103/PhysRevD.98.115017
  [arXiv:1810.01893 [hep-ph]].



\bibitem{Ferreira:2019qpf} 
  M.~M.~Ferreira, T.~B.~de Melo, S.~Kovalenko, P.~R.~D.~Pinheiro and F.~S.~Queiroz,
  Eur.\ Phys.\ J.\ C {\bf 79}, no. 11, 955 (2019)
  doi:10.1140/epjc/s10052-019-7422-z
  [arXiv:1903.07634 [hep-ph]].



\bibitem{CarcamoHernandez:2019vih} 
  A.~E.~Cárcamo Hernández, Y.~Hidalgo Velásquez and N.~A.~Pérez-Julve,
  Eur.\ Phys.\ J.\ C {\bf 79}, no. 10, 828 (2019)
  doi:10.1140/epjc/s10052-019-7325-z
  [arXiv:1905.02323 [hep-ph]].



\bibitem{CarcamoHernandez:2019lhv} 
  A.~E.~Cárcamo Hernández, D.~T.~Huong and H.~N.~Long,
  arXiv:1910.12877 [hep-ph].



\bibitem{deSousaPires:1998jc} 
  C.~A.~de Sousa Pires and O.~P.~Ravinez,
  Phys.\ Rev.\ D {\bf 58}, 035008 (1998)
  [Phys.\ Rev.\ D {\bf 58}, 35008 (1998)]
  doi:10.1103/PhysRevD.58.035008
  [hep-ph/9803409].



\bibitem{VanDong:2005ux} 
  P.~V.~Dong and H.~N.~Long,
  Int.\ J.\ Mod.\ Phys.\ A {\bf 21}, 6677 (2006)
  doi:10.1142/S0217751X06035191
  [hep-ph/0507155].



\bibitem{Montero:1998yw} 
  J.~C.~Montero, V.~Pleitez and O.~Ravinez,
  Phys.\ Rev.\ D {\bf 60}, 076003 (1999)
  doi:10.1103/PhysRevD.60.076003
  [hep-ph/9811280].



\bibitem{Montero:2005yb} 
  J.~C.~Montero, C.~C.~Nishi, V.~Pleitez, O.~Ravinez and M.~C.~Rodriguez,
  Phys.\ Rev.\ D {\bf 73}, 016003 (2006)
  doi:10.1103/PhysRevD.73.016003
  [hep-ph/0511100].



\bibitem{Pal:1994ba} 
  P.~B.~Pal,
  Phys.\ Rev.\ D {\bf 52}, 1659 (1995)
  doi:10.1103/PhysRevD.52.1659
  [hep-ph/9411406].



\bibitem{Dias:2002gg} 
  A.~G.~Dias, V.~Pleitez and M.~D.~Tonasse,
  Phys.\ Rev.\ D {\bf 67}, 095008 (2003)
  doi:10.1103/PhysRevD.67.095008
  [hep-ph/0211107].



\bibitem{Dias:2003zt} 
  A.~G.~Dias and V.~Pleitez,
  Phys.\ Rev.\ D {\bf 69}, 077702 (2004)
  doi:10.1103/PhysRevD.69.077702
  [hep-ph/0308037].



\bibitem{Dias:2003iq} 
  A.~G.~Dias, C.~A.~de S. Pires and P.~S.~Rodrigues da Silva,
  Phys.\ Rev.\ D {\bf 68}, 115009 (2003)
  doi:10.1103/PhysRevD.68.115009
  [hep-ph/0309058].



\bibitem{Mizukoshi:2010ky} 
  J.~K.~Mizukoshi, C.~A.~de S.Pires, F.~S.~Queiroz and P.~S.~Rodrigues da Silva,
  Phys.\ Rev.\ D {\bf 83}, 065024 (2011)
  doi:10.1103/PhysRevD.83.065024
  [arXiv:1010.4097 [hep-ph]].



\bibitem{Dias:2010vt} 
  A.~G.~Dias, C.~A.~de S.Pires and P.~S.~Rodrigues da Silva,
  Phys.\ Rev.\ D {\bf 82}, 035013 (2010)
  doi:10.1103/PhysRevD.82.035013
  [arXiv:1003.3260 [hep-ph]].



\bibitem{Alvares:2012qv} 
  J.~D.~Ruiz-Alvarez, C.~A.~de S.Pires, F.~S.~Queiroz, D.~Restrepo and P.~S.~Rodrigues da Silva,
  Phys.\ Rev.\ D {\bf 86}, 075011 (2012)
  doi:10.1103/PhysRevD.86.075011
  [arXiv:1206.5779 [hep-ph]].



\bibitem{Cogollo:2014jia} 
  D.~Cogollo, A.~X.~Gonzalez-Morales, F.~S.~Queiroz and P.~R.~Teles,
  JCAP {\bf 1411}, 002 (2014)
  doi:10.1088/1475-7516/2014/11/002
  [arXiv:1402.3271 [hep-ph]].



\bibitem{daSilva:2014qba} 
  P.~S.~Rodrigues da Silva,
  Phys.\ Int.\  {\bf 7}, no. 1, 15 (2016)
  doi:10.3844/pisp.2016.15.27
  [arXiv:1412.8633 [hep-ph]].



\bibitem{Altarelli:2009gn} 
  G.~Altarelli, F.~Feruglio and L.~Merlo,
  JHEP {\bf 0905}, 020 (2009)
  doi:10.1088/1126-6708/2009/05/020
  [arXiv:0903.1940 [hep-ph]].



\bibitem{Bazzocchi:2009da} 
  F.~Bazzocchi, L.~Merlo and S.~Morisi,
  Phys.\ Rev.\ D {\bf 80}, 053003 (2009)
  doi:10.1103/PhysRevD.80.053003
  [arXiv:0902.2849 [hep-ph]].



\bibitem{Bazzocchi:2009pv} 
  F.~Bazzocchi, L.~Merlo and S.~Morisi,
  Nucl.\ Phys.\ B {\bf 816}, 204 (2009)
  doi:10.1016/j.nuclphysb.2009.03.005
  [arXiv:0901.2086 [hep-ph]].



\bibitem{Toorop:2010yh} 
  R.~de Adelhart Toorop, F.~Bazzocchi and L.~Merlo,
  JHEP {\bf 1008}, 001 (2010)
  doi:10.1007/JHEP08(2010)001
  [arXiv:1003.4502 [hep-ph]].



\bibitem{Patel:2010hr} 
  K.~M.~Patel,
  Phys.\ Lett.\ B {\bf 695}, 225 (2011)
  doi:10.1016/j.physletb.2010.11.024
  [arXiv:1008.5061 [hep-ph]].



\bibitem{Morisi:2011pm} 
  S.~Morisi, K.~M.~Patel and E.~Peinado,
  Phys.\ Rev.\ D {\bf 84}, 053002 (2011)
  doi:10.1103/PhysRevD.84.053002
  [arXiv:1107.0696 [hep-ph]].



\bibitem{Altarelli:2012bn} 
  G.~Altarelli, F.~Feruglio, L.~Merlo and E.~Stamou,
  JHEP {\bf 1208}, 021 (2012)
  doi:10.1007/JHEP08(2012)021
  [arXiv:1205.4670 [hep-ph]].



\bibitem{Mohapatra:2012tb} 
  R.~N.~Mohapatra and C.~C.~Nishi,
  Phys.\ Rev.\ D {\bf 86}, 073007 (2012)
  doi:10.1103/PhysRevD.86.073007
  [arXiv:1208.2875 [hep-ph]].



\bibitem{BhupalDev:2012nm} 
  P.~S.~Bhupal Dev, B.~Dutta, R.~N.~Mohapatra and M.~Severson,
  Phys.\ Rev.\ D {\bf 86}, 035002 (2012)
  doi:10.1103/PhysRevD.86.035002
  [arXiv:1202.4012 [hep-ph]].



\bibitem{Varzielas:2012pa} 
  I.~de Medeiros Varzielas and L.~Lavoura,
  J.\ Phys.\ G {\bf 40}, 085002 (2013)
  doi:10.1088/0954-3899/40/8/085002
  [arXiv:1212.3247 [hep-ph]].



\bibitem{Ding:2013hpa} 
  G.~J.~Ding, S.~F.~King, C.~Luhn and A.~J.~Stuart,
  JHEP {\bf 1305}, 084 (2013)
  doi:10.1007/JHEP05(2013)084
  [arXiv:1303.6180 [hep-ph]].



\bibitem{Ishimori:2010fs} 
  H.~Ishimori, Y.~Shimizu, M.~Tanimoto and A.~Watanabe,
  Phys.\ Rev.\ D {\bf 83}, 033004 (2011)
  doi:10.1103/PhysRevD.83.033004
  [arXiv:1010.3805 [hep-ph]].



\bibitem{Ding:2013eca} 
  G.~J.~Ding and Y.~L.~Zhou,
  Nucl.\ Phys.\ B {\bf 876}, 418 (2013)
  doi:10.1016/j.nuclphysb.2013.08.011
  [arXiv:1304.2645 [hep-ph]].



\bibitem{Hagedorn:2011un} 
  C.~Hagedorn and M.~Serone,
  JHEP {\bf 1110}, 083 (2011)
  doi:10.1007/JHEP10(2011)083
  [arXiv:1106.4021 [hep-ph]].



\bibitem{Campos:2014zaa} 
  M.~D.~Campos, A.~E.~Cárcamo Hernández, H.~Päs and E.~Schumacher,
  Phys.\ Rev.\ D {\bf 91}, no. 11, 116011 (2015)
  doi:10.1103/PhysRevD.91.116011
  [arXiv:1408.1652 [hep-ph]].



\bibitem{VanVien:2015xha} 
  V.~V.~Vien, H.~N.~Long and D.~P.~Khoi,
  Int.\ J.\ Mod.\ Phys.\ A {\bf 30}, no. 17, 1550102 (2015)
  doi:10.1142/S0217751X1550102X
  [arXiv:1506.06063 [hep-ph]].



\bibitem{deAnda:2017yeb} 
  F.~J.~de Anda, S.~F.~King and E.~Perdomo,
  JHEP {\bf 1712}, 075 (2017)
  Erratum: [JHEP {\bf 1904}, 069 (2019)]
  doi:10.1007/JHEP12(2017)075, 10.1007/JHEP04(2019)069
  [arXiv:1710.03229 [hep-ph]].



\bibitem{deAnda:2018oik} 
  F.~J.~de Anda and S.~F.~King,
  JHEP {\bf 1807}, 057 (2018)
  doi:10.1007/JHEP07(2018)057
  [arXiv:1803.04978 [hep-ph]].



\bibitem{CarcamoHernandez:2019eme} 
  A.~E.~Cárcamo Hernández and S.~F.~King,
  Nucl.\ Phys.\ B {\bf 953}, 114950 (2020)
  doi:10.1016/j.nuclphysb.2020.114950
  [arXiv:1903.02565 [hep-ph]].



\bibitem{Chen:2019oey} 
  P.~T.~Chen, G.~J.~Ding, S.~F.~King and C.~C.~Li,
  arXiv:1906.11414 [hep-ph].



\bibitem{deMedeirosVarzielas:2019cyj} 
  I.~De Medeiros Varzielas, S.~F.~King and Y.~L.~Zhou,
  arXiv:1906.02208 [hep-ph].



\bibitem{deMedeirosVarzielas:2019hur} 
  I.~De Medeiros Varzielas, M.~Levy and Y.~L.~Zhou,
  Phys.\ Rev.\ D {\bf 100}, no. 3, 035027 (2019)
  doi:10.1103/PhysRevD.100.035027
  [arXiv:1903.10506 [hep-ph]].



\bibitem{Froggatt:1978nt} 
  C.~D.~Froggatt and H.~B.~Nielsen,
  Nucl.\ Phys.\ B {\bf 147}, 277 (1979).
  doi:10.1016/0550-3213(79)90316-X



\bibitem{Mohapatra:1986bd} 
  R.~N.~Mohapatra and J.~W.~F.~Valle,
  Phys.\ Rev.\ D {\bf 34}, 1642 (1986).
  doi:10.1103/PhysRevD.34.1642



\bibitem{Akhmedov:1995ip} 
  E.~K.~Akhmedov, M.~Lindner, E.~Schnapka and J.~W.~F.~Valle,
  Phys.\ Lett.\ B {\bf 368}, 270 (1996)
  doi:10.1016/0370-2693(95)01504-3
  [hep-ph/9507275].



\bibitem{Akhmedov:1995vm} 
  E.~K.~Akhmedov, M.~Lindner, E.~Schnapka and J.~W.~F.~Valle,
  Phys.\ Rev.\ D {\bf 53}, 2752 (1996)
  doi:10.1103/PhysRevD.53.2752
  [hep-ph/9509255].



\bibitem{Malinsky:2005bi} 
  M.~Malinsky, J.~C.~Romao and J.~W.~F.~Valle,
  Phys.\ Rev.\ Lett.\  {\bf 95}, 161801 (2005)
  doi:10.1103/PhysRevLett.95.161801
  [hep-ph/0506296].



\bibitem{Borah:2018nvu} 
  D.~Borah and B.~Karmakar,
  Phys.\ Lett.\ B {\bf 789}, 59 (2019)
  doi:10.1016/j.physletb.2018.12.006
  [arXiv:1806.10685 [hep-ph]].



\bibitem{Hirsch:2009mx} 
  M.~Hirsch, S.~Morisi and J.~W.~F.~Valle,
  Phys.\ Lett.\ B {\bf 679}, 454 (2009)
  doi:10.1016/j.physletb.2009.08.003
  [arXiv:0905.3056 [hep-ph]].



\bibitem{Dib:2014fua} 
  C.~O.~Dib, G.~R.~Moreno and N.~A.~Neill,
  Phys.\ Rev.\ D {\bf 90}, no. 11, 113003 (2014)
  doi:10.1103/PhysRevD.90.113003
  [arXiv:1409.1868 [hep-ph]].



\bibitem{Chakraborty:2014hfa} 
  M.~Chakraborty, H.~Z.~Devi and A.~Ghosal,
  Phys.\ Lett.\ B {\bf 741}, 210 (2015)
  doi:10.1016/j.physletb.2014.12.038
  [arXiv:1410.3276 [hep-ph]].



\bibitem{Sinha:2015ooa} 
  R.~Sinha, R.~Samanta and A.~Ghosal,
  Phys.\ Lett.\ B {\bf 759}, 206 (2016)
  doi:10.1016/j.physletb.2016.05.080
  [arXiv:1508.05227 [hep-ph]].



\bibitem{Salazar:2015gxa} 
  C.~Salazar, R.~H.~Benavides, W.~A.~Ponce and E.~Rojas,
  JHEP {\bf 1507}, 096 (2015)
  doi:10.1007/JHEP07(2015)096
  [arXiv:1503.03519 [hep-ph]].



\bibitem{Huyen:2012uk} 
  V.~T.~N.~Huyen, H.~N.~Long, T.~T.~Lam and V.~Q.~Phong,
  Commun.\ Phys.\  {\bf 24}, no. 2, 97 (2014)
  doi:10.15625/0868-3166/24/2/3774
  [arXiv:1210.5833 [hep-ph]].



\bibitem{Martinez:2008jj} 
  R.~Martinez and F.~Ochoa,
  Phys.\ Rev.\ D {\bf 77}, 065012 (2008)
  doi:10.1103/PhysRevD.77.065012
  [arXiv:0802.0309 [hep-ph]].



\bibitem{Buras:2013dea} 
  A.~J.~Buras, F.~De Fazio and J.~Girrbach,
  JHEP {\bf 1402}, 112 (2014)
  doi:10.1007/JHEP02(2014)112
  [arXiv:1311.6729 [hep-ph]].



\bibitem{Buras:2014yna} 
  A.~J.~Buras, F.~De Fazio and J.~Girrbach-Noe,
  JHEP {\bf 1408}, 039 (2014)
  doi:10.1007/JHEP08(2014)039
  [arXiv:1405.3850 [hep-ph]].



\bibitem{Buras:2012dp} 
  A.~J.~Buras, F.~De Fazio, J.~Girrbach and M.~V.~Carlucci,
  JHEP {\bf 1302}, 023 (2013)
  doi:10.1007/JHEP02(2013)023
  [arXiv:1211.1237 [hep-ph]].



\bibitem{Diaz:2003dk} 
  R.~A.~Diaz, R.~Martinez and F.~Ochoa,
  Phys.\ Rev.\ D {\bf 69}, 095009 (2004)
  doi:10.1103/PhysRevD.69.095009
  [hep-ph/0309280].



\bibitem{Long:2018dun} 
  H.~N.~Long, N.~V.~Hop, L.~T.~Hue, N.~H.~Thao and A.~E.~Cárcamo Hernández,
  Phys.\ Rev.\ D {\bf 100}, no. 1, 015004 (2019)
  doi:10.1103/PhysRevD.100.015004
  [arXiv:1810.00605 [hep-ph]].



\bibitem{Perez:2004jc} 
  M.~A.~Perez, G.~Tavares-Velasco and J.~J.~Toscano,
  Phys.\ Rev.\ D {\bf 69}, 115004 (2004)
  doi:10.1103/PhysRevD.69.115004
  [hep-ph/0402156].



\bibitem{Aaboud:2017sjh} 
  M.~Aaboud {\it et al.} [ATLAS Collaboration],
  JHEP {\bf 1801}, 055 (2018)
  doi:10.1007/JHEP01(2018)055
  [arXiv:1709.07242 [hep-ex]].



\bibitem{Bora:2012tx} 
  K.~Bora,
  Horizon {\bf 2}, 112 (2013)
  [arXiv:1206.5909 [hep-ph]].



\bibitem{Xing:2007fb} 
  Z.~z.~Xing, H.~Zhang and S.~Zhou,
  Phys.\ Rev.\ D {\bf 77}, 113016 (2008)
  doi:10.1103/PhysRevD.77.113016
  [arXiv:0712.1419 [hep-ph]].



\bibitem{Olive:2016xmw} 
  C.~Patrignani {\it et al.} [Particle Data Group],
  Chin.\ Phys.\ C {\bf 40}, no. 10, 100001 (2016).
  doi:10.1088/1674-1137/40/10/100001



\bibitem{Dev:2009aw} 
  P.~S.~B.~Dev and R.~N.~Mohapatra,
  Phys.\ Rev.\ D {\bf 81}, 013001 (2010)
  doi:10.1103/PhysRevD.81.013001
  [arXiv:0910.3924 [hep-ph]].



\bibitem{BhupalDev:2012zg} 
  P.~S.~Bhupal Dev, R.~Franceschini and R.~N.~Mohapatra,
  Phys.\ Rev.\ D {\bf 86}, 093010 (2012)
  doi:10.1103/PhysRevD.86.093010
  [arXiv:1207.2756 [hep-ph]].



\bibitem{Das:2012ze} 
  A.~Das and N.~Okada,
  Phys.\ Rev.\ D {\bf 88}, 113001 (2013)
  doi:10.1103/PhysRevD.88.113001
  [arXiv:1207.3734 [hep-ph]].



\bibitem{Dev:2013oxa} 
  C.~H.~Lee, P.~S.~Bhupal Dev and R.~N.~Mohapatra,
  Phys.\ Rev.\ D {\bf 88}, no. 9, 093010 (2013)
  doi:10.1103/PhysRevD.88.093010
  [arXiv:1309.0774 [hep-ph]].



\bibitem{Das:2014jxa} 
  A.~Das, P.~S.~Bhupal Dev and N.~Okada,
  Phys.\ Lett.\ B {\bf 735}, 364 (2014)
  doi:10.1016/j.physletb.2014.06.058
  [arXiv:1405.0177 [hep-ph]].



\bibitem{Das:2016hof} 
  A.~Das, P.~Konar and S.~Majhi,
  JHEP {\bf 1606}, 019 (2016)
  doi:10.1007/JHEP06(2016)019
  [arXiv:1604.00608 [hep-ph]].



\bibitem{Das:2017gke} 
  A.~Das, P.~Konar and A.~Thalapillil,
  JHEP {\bf 1802}, 083 (2018)
  doi:10.1007/JHEP02(2018)083
  [arXiv:1709.09712 [hep-ph]].



\bibitem{Das:2017nvm} 
  A.~Das and N.~Okada,
  Phys.\ Lett.\ B {\bf 774}, 32 (2017)
  doi:10.1016/j.physletb.2017.09.042
  [arXiv:1702.04668 [hep-ph]].



\bibitem{Das:2017zjc} 
  A.~Das, P.~S.~B.~Dev and C.~S.~Kim,
  Phys.\ Rev.\ D {\bf 95}, no. 11, 115013 (2017)
  doi:10.1103/PhysRevD.95.115013
  [arXiv:1704.00880 [hep-ph]].



\bibitem{Das:2017rsu} 
  A.~Das, Y.~Gao and T.~Kamon,
  Eur.\ Phys.\ J.\ C {\bf 79}, no. 5, 424 (2019)
  doi:10.1140/epjc/s10052-019-6937-7
  [arXiv:1704.00881 [hep-ph]].



\bibitem{Das:2018usr} 
  A.~Das, S.~Jana, S.~Mandal and S.~Nandi,
  Phys.\ Rev.\ D {\bf 99}, no. 5, 055030 (2019)
  doi:10.1103/PhysRevD.99.055030
  [arXiv:1811.04291 [hep-ph]].



\bibitem{Das:2018hph} 
  A.~Das,
  Adv.\ High Energy Phys.\  {\bf 2018}, 9785318 (2018)
  doi:10.1155/2018/9785318
  [arXiv:1803.10940 [hep-ph]].



\bibitem{Bhardwaj:2018lma} 
  A.~Bhardwaj, A.~Das, P.~Konar and A.~Thalapillil,
  arXiv:1801.00797 [hep-ph].



\bibitem{Helo:2018rll} 
  J.~C.~Helo, H.~Li, N.~A.~Neill, M.~Ramsey-Musolf and J.~C.~Vasquez,
  Phys.\ Rev.\ D {\bf 99}, no. 5, 055042 (2019)
  doi:10.1103/PhysRevD.99.055042
  [arXiv:1812.01630 [hep-ph]].



\bibitem{Pascoli:2018heg} 
  S.~Pascoli, R.~Ruiz and C.~Weiland,
  JHEP {\bf 1906}, 049 (2019)
  doi:10.1007/JHEP06(2019)049
  [arXiv:1812.08750 [hep-ph]].



\bibitem{Krastev:1988yu} 
  P.~I.~Krastev and S.~T.~Petcov,
  Phys.\ Lett.\ B {\bf 205}, 84 (1988).
  doi:10.1016/0370-2693(88)90404-2



\bibitem{deSalas:2017kay} 
  P.~F.~de Salas, D.~V.~Forero, C.~A.~Ternes, M.~Tortola and J.~W.~F.~Valle,
  Phys.\ Lett.\ B {\bf 782}, 633 (2018)
  doi:10.1016/j.physletb.2018.06.019
  [arXiv:1708.01186 [hep-ph]].



\bibitem{Shifman:1979eb} 
  M.~A.~Shifman, A.~I.~Vainshtein, M.~B.~Voloshin and V.~I.~Zakharov,
  Sov.\ J.\ Nucl.\ Phys.\  {\bf 30}, 711 (1979)
  [Yad.\ Fiz.\  {\bf 30}, 1368 (1979)].



\bibitem{Gavela:1981ri} 
  M.~B.~Gavela, G.~Girardi, C.~Malleville and P.~Sorba,
  Nucl.\ Phys.\ B {\bf 193}, 257 (1981).
  doi:10.1016/0550-3213(81)90529-0



\bibitem{Kalyniak:1985ct} 
  P.~Kalyniak, R.~Bates and J.~N.~Ng,
  Phys.\ Rev.\ D {\bf 33}, 755 (1986).
  doi:10.1103/PhysRevD.33.755



\bibitem{Gunion:1989we} 
  J.~F.~Gunion, H.~E.~Haber, G.~L.~Kane and S.~Dawson,
  Front.\ Phys.\  {\bf 80}, 1 (2000).



\bibitem{Spira:1997dg} 
  M.~Spira,
  Fortsch.\ Phys.\  {\bf 46}, 203 (1998)
  doi:10.1002/(SICI)1521-3978(199804)46:3<203::AID-PROP203>3.0.CO;2-4
  [hep-ph/9705337].



\bibitem{Djouadi:2005gj} 
  A.~Djouadi,
  Phys.\ Rept.\  {\bf 459}, 1 (2008)
  doi:10.1016/j.physrep.2007.10.005
  [hep-ph/0503173].



\bibitem{Marciano:2011gm} 
  W.~J.~Marciano, C.~Zhang and S.~Willenbrock,
  Phys.\ Rev.\ D {\bf 85}, 013002 (2012)
  doi:10.1103/PhysRevD.85.013002
  [arXiv:1109.5304 [hep-ph]].



\bibitem{Wang:2012gm} 
  L.~Wang and X.~F.~Han,
  Phys.\ Rev.\ D {\bf 86}, 095007 (2012)
  doi:10.1103/PhysRevD.86.095007
  [arXiv:1206.1673 [hep-ph]].



\bibitem{Carcamo-Hernandez:2013ypa} 
  A.~E.~Carcamo Hernandez, C.~O.~Dib and A.~R.~Zerwekh,
  Eur.\ Phys.\ J.\ C {\bf 74}, 2822 (2014)
  doi:10.1140/epjc/s10052-014-2822-6
  [arXiv:1304.0286 [hep-ph]].



\bibitem{Bhattacharyya:2014oka} 
  G.~Bhattacharyya and D.~Das,
  Phys.\ Rev.\ D {\bf 91}, 015005 (2015)
  doi:10.1103/PhysRevD.91.015005
  [arXiv:1408.6133 [hep-ph]].



\bibitem{Fortes:2014dia} 
  E.~C.~F.~S.~Fortes, A.~C.~B.~Machado, J.~Montaño and V.~Pleitez,
  J.\ Phys.\ G {\bf 42}, no. 11, 115001 (2015)
  doi:10.1088/0954-3899/42/11/115001
  [arXiv:1408.0780 [hep-ph]].



\bibitem{Hernandez:2015xka} 
  A.~E.~Carcamo Hernandez, C.~O.~Dib and A.~R.~Zerwekh,
  Nucl.\ Part.\ Phys.\ Proc.\  {\bf 267-269}, 35 (2015)
  doi:10.1016/j.nuclphysbps.2015.10.079
  [arXiv:1503.08472 [hep-ph]].



\bibitem{Hernandez:2015dga} 
  A.~E.~Cárcamo Hernández, I.~de Medeiros Varzielas and E.~Schumacher,
  Phys.\ Rev.\ D {\bf 93}, no. 1, 016003 (2016)
  doi:10.1103/PhysRevD.93.016003
  [arXiv:1509.02083 [hep-ph]].



\bibitem{CarcamoHernandez:2017pei} 
  A.~E.~Cárcamo Hernández, B.~Díaz Sáez, C.~O.~Dib and A.~Zerwekh,
  Phys.\ Rev.\ D {\bf 96}, no. 11, 115027 (2017)
  doi:10.1103/PhysRevD.96.115027
  [arXiv:1707.05195 [hep-ph]].



\bibitem{Khachatryan:2014ira} 
  V.~Khachatryan {\it et al.} [CMS Collaboration],
  Eur.\ Phys.\ J.\ C {\bf 74}, no. 10, 3076 (2014)
  doi:10.1140/epjc/s10052-014-3076-z
  [arXiv:1407.0558 [hep-ex]].



\bibitem{Aad:2014eha} 
  G.~Aad {\it et al.} [ATLAS Collaboration],
  Phys.\ Rev.\ D {\bf 90}, no. 11, 112015 (2014)
  doi:10.1103/PhysRevD.90.112015
  [arXiv:1408.7084 [hep-ex]].



\bibitem{Ishimori:2010au} 
  H.~Ishimori, T.~Kobayashi, H.~Ohki, Y.~Shimizu, H.~Okada and M.~Tanimoto,
  Prog.\ Theor.\ Phys.\ Suppl.\  {\bf 183}, 1 (2010)
  doi:10.1143/PTPS.183.1
  [arXiv:1003.3552 [hep-th]].




\end{thebibliography}
\end{document}